\documentclass{article}

\usepackage{arxiv}

\usepackage[utf8]{inputenc} 
\usepackage[T1]{fontenc}    
\usepackage{hyperref}       
\usepackage{url}            
\usepackage{booktabs}       
\usepackage{amsmath}        
\usepackage{amsfonts}       
\usepackage{nicefrac}       
\usepackage{microtype}      
\usepackage{cleveref}       
\usepackage{graphicx}
\usepackage{makecell}       
\usepackage{natbib}
\usepackage{doi}

\usepackage{tikz}
\usetikzlibrary{arrows.meta,positioning,calc}
\usepackage{pgfplots}
\usepgfplotslibrary{groupplots}
\pgfplotsset{compat=1.18}

\title{Practical Assessment of Agentic Systems in Acoustics}

\usepackage{authblk}

\author[1]{Arne Hildenbrand\thanks{Corresponding author: \texttt{arne.hildenbrand@tum.de}}}
\author[2]{Philipp Diel}
\author[3]{Zhuo Cao}
\author[3]{Hanno Scharr}
\author[2]{Marco Günther}
\author[2]{Stefan Pischinger}
\author[1]{Steffen Marburg}
\affil[1]{Chair of Vibroacoustics of Vehicles and Machines, Department of Engineering Physics and Computation, TUM School of Engineering and Design, Technical University of Munich, Boltzmannstrasse 15, 85748 Garching, Germany}
\affil[2]{Chair of Thermodynamics of Mobile Energy Conversion Systems, Faculty of Mechanical Engineering, RWTH Aachen University, Forckenbeckstrasse 4, 52074 Aachen, Germany}
\affil[3]{IAS-8, Forschungszentrum J\"ulich, Wilhelm-Johnen-Stra\ss{}e, 52425 J\"ulich, Germany}

\renewcommand{\shorttitle}{Practical Assessment of Agentic Systems in Acoustics}

\hypersetup{
pdftitle={Practical Assessment of Agentic Systems in Acoustics},
pdfsubject={cs.},
pdfauthor={Arne Hildenbrand, Philipp Diel, Zhuo Cao, Hanno Scharr, Stefan Pischinger, Steffen Marburg},
pdfkeywords={AI agents, large language models, agent harness, benchmark, fault detection, acoustic metamaterials},
}

\begin{document}
\maketitle

\begin{abstract}
In engineering, applications of artificial intelligence have so far been largely limited to surrogate models 
and optimization tasks. Large language models changed this, as they can write software from a task given in words. Placed inside a
harness, a system of components that lets it search for information and execute software, such a model becomes
an agent that can take over recurring engineering tasks. This work measures how the model 
and the elements of the harness shape the performance and behavior of the agent in two common acoustic applications. 
In the first application, the agent detects and locates faults in audio recordings from a vehicle interior with near-perfect 
accuracy. A benchmark on this task shows that a weak model sets a limit that no harness design can overcome, 
and that design choices take effect only once the model is capable enough to use them. In the second application, a case study 
on the design of a metamaterial unit cell, the agent reliably finds designs that achieve a specified insertion loss 
over a given frequency range. The results aim to help the reader understand what each element of an agent 
contributes and how to design an agent for acoustic tasks.
\end{abstract}

\keywords{AI agents \and Large language models \and Agent harness \and Benchmark \and Fault detection \and Acoustic metamaterials}

\section{\label{sec:1} Introduction}
In acoustics, the methods that are combined under the term artificial intelligence (AI) are mostly used as optimization tools or as surrogate models
for time consuming simulations or measurements. When applied like this they are often referred to as machine learning and are well 
established in acoustics \citep{cunhaReviewMachineLearning2023,schmidPhysicsinformedNeuralNetworks2024,biancoMachineLearningAcoustics2019}. 
Recent developments in the field of large language models (LLM)
allow artificial intelligence to interpret complex texts and to write pieces of software independently. Given a task, the model 
relies on the user for additional information and it is not able to take actions with the information that is generated by it.  
This changes when the language model is placed inside a system of components that allow the LLM e.g. to search for new information or
execute pieces of software needed to fulfill a task. This system is called a \emph{harness} and together with the LLM it builds an \emph{agent} 
\citep{fangComprehensiveSurveySelfEvolving2025,luoLargeLanguageModel2025}.
An agent can therefore write and execute software on its own, and by this take over or support recurring engineering
tasks. Ready-made pieces of software that the agent can invoke are referred to as \emph{skills}.
 
Several engineering and scientific disciplines have already built agents for that purpose
\citep{ai4scienceImpactLargeLanguage2023a}. ChemCrow for chemistry couples a language model to a
set of chemistry tools \citep{branChemCrowAugmentingLargelanguage2023a}, and a separate study measures what a model
achieves without them \citep{mirzaAreLargeLanguage2024}. HoneyComb for materials science equips an agent with a curated
knowledge base and a skill set \citep{zhangHoneyCombFlexibleLLMBased2024}, and MechGPT adapts the model itself to texts
on mechanics \citep{buehlerMechGPTLanguageBasedStrategy2024a}. In optics, an agent controls laboratory devices through a layer of tool
interfaces. On a benchmark, it reaches success rates far above a baseline that generates control code directly
\citep{wangAgenticAIScalable2026a}. In photonics, an agentic system turns a target spectrum into a metamaterial design
and trains its own surrogate model along the way \citep{luAgenticFrameworkAutonomous2025}.
 
AI agents have been tried in the field of acoustics as well. One work used an agentic framework to coordinate audio-language models as 
tools for question answering on recordings \citep{wijngaardAudioToolAgentAgenticFramework2026}. 
In machine listening, a language model with function calling
synthesizes anomalous machine sound to evaluate a detector \citep{purohitMIMIIAgentLeveragingLLMs2025}. 
An acoustic agent has also been used in a simulation setting that optimizes an aircraft wing, where it predicts the airfoil noise 
\citep{xuEngineeringaiPlatformTeams2025}.
 
Each of these works reports one finished agentic framework and shows what each agent achieves in its domain. A benchmark in
finite element analysis takes a different approach \citep{mudurFEABenchEvaluatingLanguage2025}. It has a language model
operate COMSOL Multiphysics \citep{comsolCOMSOLMultiphysics} through the programming interface of the software and grades the computed value against a
reference, so the agent is measured and not only demonstrated. 

This work adopts the approach of measuring the agent, but with a wider focus on the influence of the elements of the
harness on the performance and the overall behavior of the agent. It addresses the acoustic engineer who runs
simulations and analyses measurements and considers automating that work with an AI agent. The test therefore uses
two common applications, the detection of faults in acoustic measurements and the design of a metamaterial. The agent
is thus the subject of the study, and acoustics provides the standard against which it is measured. 

Both applications proved achievable for the agent, and the benchmark separates well what each element of the harness
contributes. The choice of the language model is examined in the same way. Beyond these results, this work collects
the vocabulary that the domain literature uses for harnesses and agentic systems, places the agent built here within
it, names the alternatives open at each decision, and turns the benchmark experience into practical advice for
designing each part. 

The paper is structured as follows. Section \ref{sec:2} gives an overview on the established vocabulary of agentic
systems, which is divided into the harness and its elements, detailing what each provides, and places the agent built
here within it. Sections \ref{sec:3} and \ref{sec:4} test the elements of the harness in two studies.
Section \ref{sec:3} builds the benchmark around the detection of faults in acoustic measurements. 
Section \ref{sec:4} applies the same agent to the design of a metamaterial. Section \ref{sec:5} concludes this study
and explores future work aspects.
\section{\label{sec:2} AI Agents}
This section introduces the terminology employed in the field of AI agents. It provides the vocabulary necessary to examine a given
application and to determine which of the available options is suitable for it. It also serves as the foundation for the
design decisions of this work, which are stated in Section \ref{subsec:2:3}.

The starting point is the agent itself. An agent combines a foundational model, in most cases an LLM
\citep{fangComprehensiveSurveySelfEvolving2025,sapkotaAIAgentsVs2026,kadekodiAgentFluxDecoupledFineTuning2025}, with a
surrounding framework called the agent harness.
The LLM acts as the central processing unit of the agent.
As coordinator, it interprets its input, draws conclusions, and decides on the next steps, while the harness provides
information and executes the decisions of the LLM
\citep{fangComprehensiveSurveySelfEvolving2025,tranMultiAgentCollaborationMechanisms2025,kadekodiAgentFluxDecoupledFineTuning2025}.
\subsection{\label{subsec:2:1} Agent Harness}
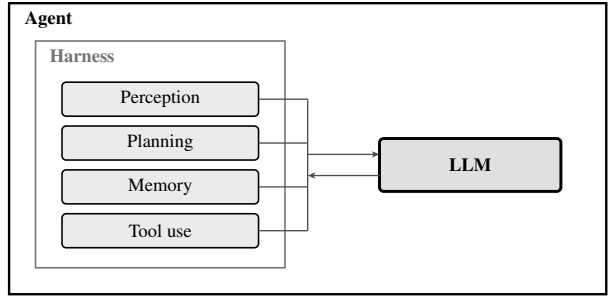
\begin{figure}
\centering
\tikzset{
  fmmod/.style   ={draw=black, line width=0.6pt, fill=black!8, rounded corners=1.6pt,
                   minimum width=2.6cm, minimum height=0.45cm, inner sep=0pt,
                   font=\scriptsize, text=black},
  fmcore/.style  ={draw=black, line width=1.1pt, fill=black!12, rounded corners=1.6pt,
                   minimum width=2.4cm, minimum height=0.70cm, inner sep=0pt,
                   font=\scriptsize\bfseries, text=black},
  fmagent/.style ={draw=black, line width=0.9pt, sharp corners},
  fmharn/.style  ={draw=black!55, line width=0.6pt, sharp corners},
  fmbus/.style   ={draw=black!70, line width=0.5pt},
  fmar/.style    ={draw=black!70, line width=0.5pt,
                   -{Stealth[length=3pt,width=2.4pt]}},
  fmhead/.style  ={font=\scriptsize\bfseries, text=black, anchor=north west,
                   inner sep=0pt},
  fmsub/.style   ={font=\scriptsize\bfseries, text=black!55, anchor=north west,
                   inner sep=0pt},
}
\begin{tikzpicture}[x=1cm, y=1cm]
\draw[fmagent] (0.05, 0.10) rectangle (7.95, 3.95);
\node[fmhead] at (0.25, 3.85) {Agent};
\draw[fmharn]  (0.40, 0.45) rectangle (3.70, 3.45);
\node[fmsub] at (0.58, 3.35) {Harness};
\node[fmmod] (pe) at (2.05, 2.68) {Perception};
\node[fmmod] (pl) at (2.05, 2.10) {Planning};
\node[fmmod] (me) at (2.05, 1.52) {Memory};
\node[fmmod] (tu) at (2.05, 0.94) {Tool use};
\node[fmcore] (llm) at (6.15, 1.81) {LLM};
\foreach \y in {2.68, 2.10, 1.52, 0.94} \draw[fmbus] (3.35, \y) -- (4.00, \y);
\draw[fmbus] (4.00, 0.94) -- (4.00, 2.68);
\draw[fmar] (4.00, 1.95) -- (4.95, 1.95);
\draw[fmar] (4.95, 1.67) -- (4.00, 1.67);
\end{tikzpicture}
\caption{\label{fig:modules}The LLM and the harness around it form the agent. The functions the harness provides are
grouped into four modules.}
\end{figure}

The foundational model establishes the upper limit of what an agent can achieve and exerts the greatest single influence
on its performance \citep{plaatAgenticLargeLanguage2025,xiRisePotentialLarge2023}. The harness determines how
closely an agent approaches this limit \citep{fanHarnessMUSafeGoverned2026}. For a practitioner, the harness is the part that can be freely designed. Its scope ranges from a thin wrapper around a single LLM
invocation to a complete system infrastructure \citep{guModelScalingSystem2026}. Across this range, the functions a
harness must provide remain constant. They are commonly grouped into four modules, namely perception, planning, memory,
and tool use
\citep{fangComprehensiveSurveySelfEvolving2025,wangSurveyLargeLanguage2024,weiArchitecturalDesignDecisions2026}, as
shown in Figure \ref{fig:modules}.
Of these modules, tool use is the most relevant for this work. Expert knowledge enters the agent through this module in
the form of skills. A skill is a self-contained executable unit the agent can invoke, for instance a Python script. The
planning module organizes
the work within a multi-agent system (MAS) \citep{fangComprehensiveSurveySelfEvolving2025}, which is broadly referred to
as an agentic system \citep{sapkotaAIAgentsVs2026,tranMultiAgentCollaborationMechanisms2025}. This is the structure of the agent used here.
Perception describes the
manner in which the agent processes its material and falls outside the scope of this work. Memory becomes significant in
extended conversations, in which the agent must maintain context and retrieve information from prior sessions to remain
consistent. As each test in this work comprises a single user message, the mechanisms for memory storage and retrieval
are not examined and beyond the scope of this work.

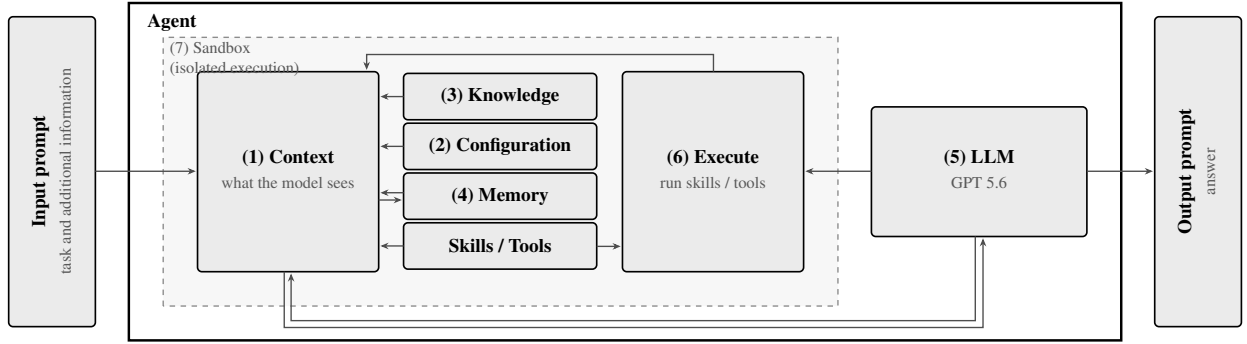
\begin{figure}
\resizebox{\textwidth}{!}{
\tikzset{
  hobox/.style   ={draw=black, line width=0.6pt, fill=black!8, rounded corners=1.6pt},
  hoouter/.style ={draw=black, line width=0.9pt, sharp corners},
  hosand/.style  ={draw=black!45, line width=0.5pt, dash pattern=on 2pt off 2pt,
                   fill=black!3, sharp corners},
  hoar/.style    ={draw=black!70, line width=0.5pt,
                   -{Stealth[length=3pt,width=2.4pt]}},
  hohead/.style  ={font=\scriptsize\bfseries, text=black, inner sep=0pt},
  hosub/.style   ={font=\tiny, text=black!65, inner sep=0pt},
}
\begin{tikzpicture}[x=\dimexpr\textwidth/15\relax, y=\dimexpr\textwidth/15\relax]
\draw[hoouter] (1.69,0.18) rectangle (13.25,4.11);
\draw[hosand]  (2.09,0.58) rectangle (9.95,3.71);
\draw[hobox]   (2.49,0.98) rectangle (4.60,3.31);
\draw[hobox]   (4.89,2.75) rectangle (7.14,3.29);
\draw[hobox]   (4.89,2.17) rectangle (7.14,2.71);
\draw[hobox]   (4.89,1.59) rectangle (7.14,2.13);
\draw[hobox]   (4.89,1.01) rectangle (7.14,1.55);
\draw[hobox]   (7.44,0.98) rectangle (9.55,3.31);
\draw[hobox]   (10.34,1.39) rectangle (12.85,2.90);
\draw[hobox]   (0.29,0.34) rectangle (1.30,3.95);
\draw[hobox]   (13.64,0.34) rectangle (14.65,3.95);
\node[hohead, anchor=west] at (1.90,3.88) {Agent};
\node[hosub, anchor=north west, align=left] at (2.16,3.66) {(7) Sandbox\\(isolated execution)};
\node[hohead] at (3.545,2.30) {(1) Context};
\node[hosub]  at (3.545,2.02) {what the model sees};
\node[hohead] at (6.015,3.02) {(3) Knowledge};
\node[hohead] at (6.015,2.44) {(2) Configuration};
\node[hohead] at (6.015,1.86) {(4) Memory};
\node[hohead] at (6.015,1.28) {Skills / Tools};
\node[hohead] at (8.495,2.30) {(6) Execute};
\node[hosub]  at (8.495,2.02) {run skills / tools};
\node[hohead] at (11.595,2.30) {(5) LLM};
\node[hosub]  at (11.595,2.02) {GPT 5.6};
\node[hohead, rotate=90] at (0.66,2.15) {Input prompt};
\node[hosub,  rotate=90] at (0.95,2.15) {task and additional information};
\node[hohead, rotate=90] at (14.01,2.15) {Output prompt};
\node[hosub,  rotate=90] at (14.30,2.15) {answer};
\draw[hoar] (1.30,2.15)  -- (2.47,2.15);
\draw[hoar] (12.85,2.15) -- (13.62,2.15);
\draw[hoar] (4.89,3.02)  -- (4.62,3.02);
\draw[hoar] (4.89,2.44)  -- (4.62,2.44);
\draw[hoar] (4.89,1.90)  -- (4.62,1.90);
\draw[hoar] (4.60,1.815) -- (4.87,1.815);
\draw[hoar] (4.89,1.28)  -- (4.62,1.28);
\draw[hoar] (7.14,1.275) -- (7.42,1.275);
\draw[hoar] (10.34,2.15) -- (9.57,2.15);
\draw[hoar] (8.50,3.31) -- (8.50,3.51) -- (4.46,3.51) -- (4.46,3.33);
\draw[hoar] (3.50,0.98) -- (3.50,0.33) -- (11.64,0.33) -- (11.64,1.37);
\draw[hoar] (11.55,1.39) -- (11.55,0.41) -- (3.58,0.41) -- (3.58,0.96);
\end{tikzpicture}}
\caption{\label{fig:harness}The elements of an agent harness from an implementation perspective. The prompt enters the context, together with the mounted knowledge, 
the configuration files, the memory and the skills, all assembled by the harness. The model reasons over
this context and decides which skills to execute. The result of the execution re-enters the context, and the loop repeats until an
answer is returned. The numbers follow the order in which Section \ref{subsec:2:1} describes the elements. Everything inside the dashed boundary runs in an isolated execution environment.}
\end{figure}

These four modules, perception, planning, memory and tool use identify the functions that an agent harness fulfills. For practitioners, a more detailed
breakdown is advantageous, as it describes which elements provide these functions in practice and how they are connected
to form a harness. Knowing the contribution of each element allows the reader to retrace the decisions made for the
agent in this work and to judge which elements to adjust for their own application. Figure \ref{fig:harness} shows these
elements and the loop they form, which the remainder of this section discusses.
The loop that connects these elements is the one described by ReAct \citep{yaoReActSynergizingReasoning2023}. The model
reasons over the context, calls a tool, and reads the result before reasoning again. As a result, each turn consists of
a reasoning phase and an action phase. The following paragraphs describe the elements in the order in which the loop 
passes them in Figure \ref{fig:harness}.

\emph{(1) Context.} The prompt is passed to the agent and becomes an integral component of the context, which encompasses everything the LLM receives during a single
invocation. The harness assembles the context at each step of the loop, and the context window of the model limits its
size. Information not contained within the context is inaccessible to the model. In addition to the input prompt, the
context holds the \emph{configuration} of the agent, the \emph{knowledge} mounted into the workspace, the \emph{memory}
of what the agent has generated in the current or previous sessions, and the \emph{skills} available for execution (see Figure \ref{fig:harness}). 

\emph{(2) Configuration.} The configuration is a set of files containing the operating instructions of the agent. Its most consequential entry is 
the catalog of available skills and instructions on how to use them. Placing a skill in the workspace of the agent is 
not sufficient for the agent to use it. The skill also has to be documented and announced to the agent, 
and the way this is done is a design decision of the engineer.

\emph{(3) Knowledge.} Knowledge is defined as the documents and other information sources about its domain that are made available to an
agent, as distinct from the memory the agent builds up from its own experience \citep{luoStorageExperienceSurvey2026}.
Knowledge reaches the model by three routes. In fine-tuning, the knowledge is written into the model weights by training
on a collection of domain documents, a knowledge archive, so that it becomes internal knowledge of the LLM
\citep{wangSurveyLargeLanguage2024}. This route has been taken in acoustics, where a model was trained on a dataset of
metamaterial geometries and their sound absorption coefficients and then used both to predict the absorption of a given
geometry and to propose a geometry for a target spectrum \citep{jiangDatadrivenDesignSound2025}. In retrieval-augmented
generation, a retrieval module indexes an external archive, selects the passages that match the current prompt, and
inserts them into the context without the involvement of the agent
\citep{lewisRetrievalAugmentedGenerationKnowledgeIntensive2021,gaoRetrievalAugmentedGenerationLarge2024,fangComprehensiveSurveySelfEvolving2025}.
In the third route, the agent retrieves the knowledge itself as needed
\citep{wangSurveyLargeLanguage2024,wangMCPBenchBenchmarkingToolUsing2025,xiRisePotentialLarge2023}. The
routes differ in how deeply the knowledge is embedded and in where the decision on its use is made. Knowledge stored in
the weights is part of the LLM and can only be changed by retraining. Knowledge provided by a retrieval module stays
outside the model and is placed into the context by the harness in every run. This is an explicit design decision, since
the archive and the selection mechanism are fixed when the harness is built. Knowledge retrieved by the agent also stays
outside the model, but it enters the context only if the agent decides to read it.

\emph{(4) Memory.} Memory holds the information an agent retains beyond a single model invocation. The literature distinguishes between
short-term or working memory, which stores information within the current context of a session, and long-term memory,
which persists across sessions and is retrieved as needed
\citep{luoStorageExperienceSurvey2026,xiRisePotentialLarge2023,wangSurveyLargeLanguage2024}. Short-term memory is
limited to the session.
Long-term memory is built up in three stages, from storage through reflection to experience. The agent records the
course of a session, meaning the sequence of steps it performs, condenses these records, and abstracts them into
experience that it can apply to new tasks \citep{luoStorageExperienceSurvey2026}. Keeping this layer trustworthy is one
of the central bottlenecks of a harness \citep{guModelScalingSystem2026}. Stored information must remain accurate over
time, or the agent's own past becomes a source of error.

\emph{(5) LLM (Reasoning).} Once assembled, the context is transferred to the model, and the processing within the model prior to generating an
answer is its reasoning. The reasoning effort, also known as test-time computation, can be adjusted by the user
and current models make it accessible as a setting \citep{snellScalingLLMTestTime2024,openaiLearningReasonLLMs2024}.
The additional computation costs more, as the model generates more intermediate
steps before it acts, which increases the latency of a run
\citep{kadekodiAgentFluxDecoupledFineTuning2025,weiChainofThoughtPromptingElicits2023}. Each reasoning step ends in a
decision and the next action of the agent, which is executed by the harness.

\emph{(6) Execution.} This action is the second phase of the ReAct loop and consists of the execution of a skill. The result of this invocation 
re-enters the context, and the loop continues. The harness provides the fundamental tools, such as an execution shell or 
a Python interpreter (Python REPL \citep{zhangHoneyCombFlexibleLLMBased2024}). Existing agent frameworks already include 
these tools, so only the domain skills have to be added. 

\emph{(7) Sandbox (Isolation).} The environment is isolated in a container, which is a packaged filesystem that runs isolated from the host
system, as illustrated in Figure \ref{fig:harness}. The isolation serves two purposes. It confines the agent to the
content inside the boundary, and it resets every run to the same initial state, which makes the benchmark runs
comparable.

The elements described above vary in their accessibility to a domain expert. The composition of the context and the
design of the loop are properties of the harness, and modifying them necessitates altering the harness itself. Harness
or loop engineering is a powerful discipline of its own and is beyond the focus of this work. Domain knowledge,
agent configuration, and skills are accessible without engaging in this discipline. These elements are provided around an
existing harness and require expertise only in the respective field, which is why this work focuses on them.
\subsection{\label{subsec:2:2} Agentic Systems}
\begin{figure}
\colorlet{msLine}{black}
\colorlet{msOuter}{black}
\colorlet{msFill}{black!8}
\colorlet{msArrow}{black!70}
\tikzset{
  msnode/.style  ={circle, draw=msLine, line width=0.6pt, fill=msFill,
                   minimum size=3.0mm, inner sep=0pt},
  mshub/.style   ={circle, draw=msOuter, line width=0.6pt, fill=msOuter,
                   minimum size=3.6mm, inner sep=0pt},
  msrole/.style  ={circle, draw=msLine, line width=0.6pt, fill=msFill,
                   minimum size=4.0mm, inner sep=0pt},
  msrule/.style  ={draw=msLine, line width=0.6pt, fill=msFill, rounded corners=0.5pt,
                   minimum width=3.4mm, minimum height=2.6mm, inner sep=0pt,
                   text=msOuter, font=\tiny},
  msedge/.style  ={draw=msArrow, line width=0.5pt},
  msweak/.style  ={draw=black!35, line width=0.5pt},
  msar/.style    ={draw=msArrow, line width=0.5pt,
                   -{Stealth[length=3pt,width=2.4pt]}},
  msbranch/.style={draw=black!45, line width=0.5pt,
                   dash pattern=on 1.2pt off 1.2pt,
                   -{Stealth[length=2.6pt,width=2.2pt]}},
  msframe/.style ={draw=black!45, line width=0.5pt, rounded corners=1.4pt},
  mslab/.style   ={font=\scriptsize, text=msOuter, anchor=north, inner sep=0pt},
  msnum/.style   ={font=\tiny, text=msOuter, inner sep=0pt},
  mslegend/.style={font=\scriptsize, text=msOuter, anchor=west, inner sep=0pt},
  mshead/.style  ={font=\scriptsize\bfseries, text=msOuter, anchor=north west,
                   inner sep=0pt},
}
\newcommand{\msgoalat}[1]{%
  \draw[draw=msOuter, fill=msArrow, line width=0.5pt]
    ([yshift=1.8mm]#1) -- ([xshift=1.8mm]#1) -- ([yshift=-1.8mm]#1) -- ([xshift=-1.8mm]#1) -- cycle;}
\begin{tikzpicture}[x=1cm, y=1cm]
\begin{scope}[yshift=0cm]
\node[mshead] at (0.000, 1.45) {Communication structure};
\node[mshub]  (h)  at (0.950, 0.000) {};
\node[msnode] (s1) at (0.950, 0.780) {};
\node[msnode] (s2) at (0.208, 0.241) {};
\node[msnode] (s3) at (0.491,-0.631) {};
\node[msnode] (s4) at (1.409,-0.631) {};
\node[msnode] (s5) at (1.692, 0.241) {};
\foreach \n in {s1,s2,s3,s4,s5} \draw[msedge] (h) -- (\n);
\node[mslab] at (0.950,-1.10) {centralized};
\node[msnode] (d1) at (5.722,0.780) {};
\node[msnode] (d2) at (5.047,0.390) {};
\node[msnode] (d3) at (5.047,-0.390) {};
\node[msnode] (d4) at (5.722,-0.780) {};
\node[msnode] (d5) at (6.397,-0.390) {};
\node[msnode] (d6) at (6.397,0.390) {};
\draw[msedge] (d1) -- (d2) -- (d3) -- (d4) -- (d5) -- (d6) -- (d1);
\draw[msedge] (d1) -- (d3);
\draw[msedge] (d4) -- (d6);
\node[mslab] at (5.722,-1.10) {decentralized};
\node[mshub]  (t)  at (11.091,0.740) {};
\node[msnode] (mL) at (10.711,0.000) {};
\node[msnode] (mR) at (11.471,0.000) {};
\node[msnode] (b1) at (10.391,-0.740) {};
\node[msnode] (b2) at (10.871,-0.740) {};
\node[msnode] (b3) at (11.311,-0.740) {};
\node[msnode] (b4) at (11.791,-0.740) {};
\draw[msedge] (t) -- (mL);   \draw[msedge] (t) -- (mR);
\draw[msedge] (mL) -- (b1);  \draw[msedge] (mL) -- (b2);
\draw[msedge] (mR) -- (b3);  \draw[msedge] (mR) -- (b4);
\draw[msweak] (mL) -- (mR);
\draw[msweak] (b1) -- (b2);  \draw[msweak] (b3) -- (b4);
\node[mslab] at (11.091,-1.10) {hierarchical};
\end{scope}
\begin{scope}[yshift=-3.07cm]
\node[mshead] at (0.000, 1.30) {Type of collaboration};
\coordinate (g1) at (0.950, 0.620);
\node[msnode] (c1) at (0.400,-0.420) {};
\node[msnode] (c2) at (1.500,-0.420) {};
\msgoalat{g1}
\draw[msedge] (c1) -- (c2);
\draw[msar, shorten >=5pt] (0.950,-0.420) -- (g1);
\node[mslab] at (0.950,-0.88) {cooperation};
\coordinate (g2) at (5.722,0.620);
\node[msnode] (k1) at (5.172,-0.420) {};
\node[msnode] (k2) at (6.272,-0.420) {};
\msgoalat{g2}
\draw[msar, shorten >=5pt] (k1) -- (g2);
\draw[msar, shorten >=5pt] (k2) -- (g2);
\node[msnum, anchor=west] at (6.112,0.100) {won};
\node[mslab] at (5.722,-0.88) {competition};
\coordinate (g3) at (11.091,0.620);
\node[msnode] (q1) at (10.541,-0.420) {};
\node[msnode] (q2) at (11.641,-0.420) {};
\msgoalat{g3}
\draw[msedge] (q1) -- (q2);
\draw[msar, shorten >=5pt] (q1) -- (g3);
\draw[msar, shorten >=5pt] (q2) -- (g3);
\node[msnum, anchor=east] at (10.701,0.100) {won};
\node[mslab] at (11.091,-0.88) {coopetition};
\end{scope}
\begin{scope}[yshift=-5.93cm]
\node[mshead] at (0.000, 1.38) {Collaboration strategy};
\node[msnode] (r1) at (0.950, 0.720) {};
\node[msnode] (r2) at (0.326,-0.360) {};
\node[msnode] (r3) at (1.574,-0.360) {};
\draw[msedge] (r1) -- (r2);  \draw[msedge] (r2) -- (r3);  \draw[msedge] (r3) -- (r1);
\node[msrule, rotate=60]  at (0.638, 0.180) {$\equiv$};
\node[msrule]             at (0.950,-0.360) {$\equiv$};
\node[msrule, rotate=-60] at (1.262, 0.180) {$\equiv$};
\node[mslab] at (0.950,-0.78) {rule based};
\node[msrole] (o1) at (5.097,0.280) {};
\node[msrole] (o2) at (5.767,0.420) {};
\node[msrole] (o3) at (5.527,-0.320) {};
\node[msrole] (o4) at (6.347,-0.100) {};
\draw[msedge] (o1) -- (o2);  \draw[msedge] (o2) -- (o3);
\draw[msedge] (o1) -- (o3);  \draw[msedge] (o2) -- (o4);
\fill[msOuter] (5.097,0.280) circle (0.052);
\fill[msOuter] (5.707,0.396) rectangle (5.827,0.444);
\fill[msOuter] (5.527,-0.260) -- (5.58,-0.354) -- (5.474,-0.354) -- cycle;
\draw[msOuter, line width=0.55pt] (6.296,-0.151) -- (6.398,-0.049);
\draw[msOuter, line width=0.55pt] (6.296,-0.049) -- (6.398,-0.151);
\node[mslab] at (5.722,-0.78) {role based};
\coordinate (g4) at (11.841,0.000);
\node[msnode] (m0) at (10.341,0.000) {};
\msgoalat{g4}
\draw[msbranch, shorten >=5pt] (m0) to[bend left=32]  (g4);
\draw[msbranch, shorten >=5pt] (m0) to[bend right=32] (g4);
\draw[msar, shorten >=5pt]     (m0) -- (g4);
\node[mslab] at (11.091,-0.78) {model based};
\end{scope}
\node[mshub]    (zA) at (15.47,1.20) {};
\node[mslegend] (lA) at (15.8,1.20) {main agent};
\node[msnode]   (zB) at (15.47,0.68) {};
\node[mslegend] (lB) at (15.8,0.68) {(sub-)agent};
\draw[msedge]        (15.26,0.16) -- (15.68,0.16);
\node[mslegend] (lC) at (15.8,0.16) {channel};
\draw[msar]          (15.26,-0.36) -- (15.68,-0.36);
\node[mslegend] (lD) at (15.8,-0.36) {reach for a goal};
\draw[msbranch]      (15.26,-0.88) -- (15.68,-0.88);
\node[mslegend] (lE) at (15.8,-0.88) {unused path};
\coordinate     (zF) at (15.47,-1.40);
\msgoalat{zF}
\node[mslegend] (lF) at (15.8,-1.40) {goal};
\node[msrule]   (zG) at (15.47,-1.92) {$\equiv$};
\node[mslegend] (lG) at (15.8,-1.92) {rule};
\fill[msOuter] (15.355,-2.330) circle (0.055);
\fill[msOuter] (15.525,-2.360) rectangle (15.645,-2.300);
\fill[msOuter] (15.355,-2.500) -- (15.415,-2.620) -- (15.295,-2.620) -- cycle;
\draw[msOuter, line width=0.55pt] (15.525,-2.620) -- (15.645,-2.500);
\draw[msOuter, line width=0.55pt] (15.525,-2.500) -- (15.645,-2.620);
\node[mslegend] (lH) at (15.8,-2.44) {role};
\draw[msframe] (15.1,1.50) rectangle (17.85,-2.74);
\end{tikzpicture}
\caption{\label{fig:masstructures}The three axes along which an agentic system is designed. 
Communication runs through one central agent, decentralized with each agent communicating with every
other agent, or hierarchical, with agents communicating within a level and with the adjacent level.
Collaboration is cooperative when the agents work together on a task, competitive when they compete
for it, and coopetitive when both happen at once. The strategy is a set of rules that decides when agents
are allowed to communicate with one another, an assignment of roles, or model based, in which case several approaches are tried and one
of them is selected.}
\end{figure}
Several autonomous agents can be combined by a loop of their own, just as the elements of a single agent are. 
As introduced in Section \ref{subsec:2:1}, the literature calls such a system a multi-agent system
\citep{fangComprehensiveSurveySelfEvolving2025,tranMultiAgentCollaborationMechanisms2025} or agentic AI
\citep{sapkotaAIAgentsVs2026}.

As with an agent, an agentic system is defined by a loop, which in this case encompasses the agents. 
This orchestration loop assigns tasks, determines the information received by each agent, and
decides when a result is accepted \citep{guModelScalingSystem2026,sapkotaLangChainVsLangGraph2025a}. The roles and
boundaries of the agents and the manner in which they interact are thus central design decisions.

The literature characterizes the design of agentic systems along three axes: the communication structure, the
type of collaboration, and the collaboration strategy
\citep{tranMultiAgentCollaborationMechanisms2025,fangComprehensiveSurveySelfEvolving2025}. Figure
\ref{fig:masstructures} shows the options on each axis.
Communication can be centralized, with all agents communicating through a single node that manages and coordinates them,
or it can be distributed among the agents in a decentralized or hierarchical structure. Collaboration is cooperative
when the agents align their goals toward a common objective, and it can also be competitive or a combination of both,
termed coopetition. The collaboration strategy determines how interactions are governed. A collaboration strategy is 
rule-based if interactions are determined by predefined rules, model-based if agents make decisions under uncertainty by anticipating probable
outcomes and role-based if work is divided among predefined roles
\citep{tranMultiAgentCollaborationMechanisms2025,fangComprehensiveSurveySelfEvolving2025}.

The strength of a multi-agent system derives from its scalability, as a group of agents can collaborate to accomplish
tasks that exceed the capacity of a single agent
\citep{tranMultiAgentCollaborationMechanisms2025,sapkotaAIAgentsVs2026}. This capacity rests in part on the separation
of agents, with each agent operating within
its own context, which contains only what its specific task requires. Separation comes at the price of coordination. The
content of one agent's context is not accessible to the other agents and must be transferred when required. These
handovers are exchanged as text \citep{wuAutoGenEnablingNextGen2023}, which can lead to misunderstandings or omissions.
An early error can propagate to subsequent agents, and the behavior arising from multiple interacting agents is less
predictable than that of a single agent
\citep{xiRisePotentialLarge2023,plaatAgenticLargeLanguage2025,tranMultiAgentCollaborationMechanisms2025}. A multi-agent
system is thus not an improvement in itself. A single
agent that solves a task is preferable to a multi-agent system that solves it equally well, and the topology should be
expanded only where the task requires it \citep{sapkotaAIAgentsVs2026,tranMultiAgentCollaborationMechanisms2025,xiRisePotentialLarge2023}.
\subsection{\label{subsec:2:3} Agent Setup}
The agentic system used in this work is based on OpenClaw, an open-source framework for agent infrastructure
\citep{openclaw}. It provides all of the elements described above, including the execution tool through which each
skill is invoked, so none of these elements require dedicated implementation. Frameworks such as LangChain expose the loop and the
remaining parts of the harness at a deeper level and allow their modification \citep{sapkotaLangChainVsLangGraph2025a}, a
depth that this work does not require. The models employed in this work are hosted externally and belong to the GPT 5.6
family. 

\begin{figure}
\tikzset{
  agbox/.style   ={draw=black, line width=0.6pt, fill=black!8, rounded corners=1.6pt},
  agouter/.style ={draw=black, line width=0.9pt, sharp corners},
  agsand/.style  ={draw=black!45, line width=0.5pt, dash pattern=on 2pt off 2pt,
                   fill=black!3, sharp corners},
  agar/.style    ={draw=black!70, line width=0.5pt,
                   -{Stealth[length=3pt,width=2.4pt]}},
  agwire/.style  ={draw=black!70, line width=0.5pt},
  aghead/.style  ={font=\scriptsize\bfseries, text=black, inner sep=0pt},
  agsub/.style   ={font=\tiny, text=black!65, inner sep=0pt},
}
\newcommand{\agentouter}{\draw[agouter] (1.69,0.18) rectangle (13.25,4.11);}
\newcommand{\agentsandbox}{\draw[agsand] (2.09,0.58) rectangle (9.95,3.71);}
\newcommand{\agentboxes}{%
  \draw[agbox]   (2.49,0.98) rectangle (4.60,3.31);
  \draw[agbox]   (4.99,2.75) rectangle (7.04,3.29);
  \draw[agbox]   (4.99,2.17) rectangle (7.04,2.71);
  \draw[agbox]   (4.99,1.59) rectangle (7.04,2.13);
  \draw[agbox]   (4.99,1.01) rectangle (7.04,1.55);
  \draw[agbox]   (7.44,0.98) rectangle (9.55,3.31);
  \draw[agbox]   (10.34,1.39) rectangle (12.85,2.90);
}
\newcommand{\agentloop}{%
  \draw[agar] (4.99,3.02)  -- (4.62,3.02);
  \draw[agar] (4.99,2.44)  -- (4.62,2.44);
  \draw[agar] (4.99,1.90)  -- (4.62,1.90);
  \draw[agar] (4.60,1.815) -- (4.97,1.815);
  \draw[agar] (4.99,1.28)  -- (4.62,1.28);
  \draw[agar] (7.04,1.275) -- (7.42,1.275);
  \draw[agar] (10.34,2.15) -- (9.57,2.15);
  \draw[agar] (8.50,3.31) -- (8.50,3.51) -- (4.46,3.51) -- (4.46,3.33);
  \draw[agar] (3.50,0.98) -- (3.50,0.33) -- (11.64,0.33) -- (11.64,1.37);
  \draw[agar] (11.55,1.39) -- (11.55,0.41) -- (3.58,0.41) -- (3.58,0.96);
}
\newcommand{\agentlabels}{%
  \node[aghead, anchor=west] at (1.90,3.88) {Agent};
  \node[agsub, anchor=north west, align=left] at (2.16,3.66) {Sandbox\\(isolated execution)};
  \node[aghead] at (3.545,2.30) {Context};
  \node[agsub]  at (3.545,2.02) {what the model sees};
  \node[aghead] at (6.015,3.02) {Knowledge};
  \node[aghead] at (6.015,2.44) {Configuration};
  \node[aghead] at (6.015,1.86) {Memory};
  \node[aghead] at (6.015,1.28) {Skills / Tools};
  \node[aghead] at (8.495,2.30) {Execute};
  \node[agsub]  at (8.495,2.02) {run skills / tools};
  \node[aghead] at (11.595,2.30) {LLM};
  \node[agsub]  at (11.595,2.02) {GPT 5.6};
}
\begin{tikzpicture}[x=\dimexpr\textwidth/15\relax, y=\dimexpr\textwidth/15\relax]
\begin{scope}
  \agentouter
  \agentsandbox
  \agentboxes
  \agentloop
  \agentlabels
  \draw[agbox] (0.29,0.34) rectangle (1.30,3.95);
  \draw[agbox] (13.64,0.34) rectangle (14.65,3.95);
  \node[aghead, rotate=90] at (0.66,2.15) {Input prompt};
  \node[agsub,  rotate=90] at (0.95,2.15) {task and additional information};
  \node[aghead, rotate=90] at (14.01,2.15) {Output prompt};
  \node[agsub,  rotate=90] at (14.30,2.15) {answer};
  \draw[agar] (1.30,2.15)  -- (2.47,2.15);
  \draw[agar] (12.85,2.15) -- (13.62,2.15);
\end{scope}
\foreach \X in {3.20, 8.27} {
  \begin{scope}[shift={(\X-0.507,-2.304)}, scale=0.30]
    \agentouter
    \agentboxes
  \end{scope}
  \node[agsub, anchor=south, text=black] at (\X+1.734,-1.05) {Subagent};
}
\draw[agwire] (8.50,0.98) -- (8.50,0.46)
  arc[start angle=90, end angle=-90, radius=0.09] -- (8.50,-0.60);
\draw[agwire] (2.88,-0.60) -- (8.50,-0.60);
\foreach \X in {3.20, 8.27} {
  \draw[agar] (\X-0.32,-0.60) -- (\X-0.32,-1.659) -- (\X+0.22,-1.659);
}
\foreach \X in {3.20, 8.27} {
  \draw[agwire] (\X+3.348,-1.659) -- (\X+3.60,-1.659) -- (\X+3.60,-2.60);
}
\draw[agar] (11.87,-2.60) -- (1.46,-2.60) -- (1.46,1.20) -- (2.47,1.20);
\node[agsub, anchor=south west] at (3.20,-0.57) {delegated task};
\node[agsub, anchor=south west] at (3.20,-2.57) {generated answer};
\end{tikzpicture}
\caption{\label{fig:agentic}A multi-agent system built from the blocks and the structure of the harness of
Figure \ref{fig:harness}. The system of Section \ref{subsec:2:3} is centralized, cooperative and role based. One
orchestrating agent spawns two subagents via its execute module, and each subagent repeats the same structure,
depicted here as a pictogram. The orchestration agent writes the task into the context of a subagent, and every
subagent keeps its own context, its own harness and its own calls to the model. Only the final message of a
subagent returns into the context of the orchestrating agent, so the subagents never communicate
with one another.}
\end{figure}
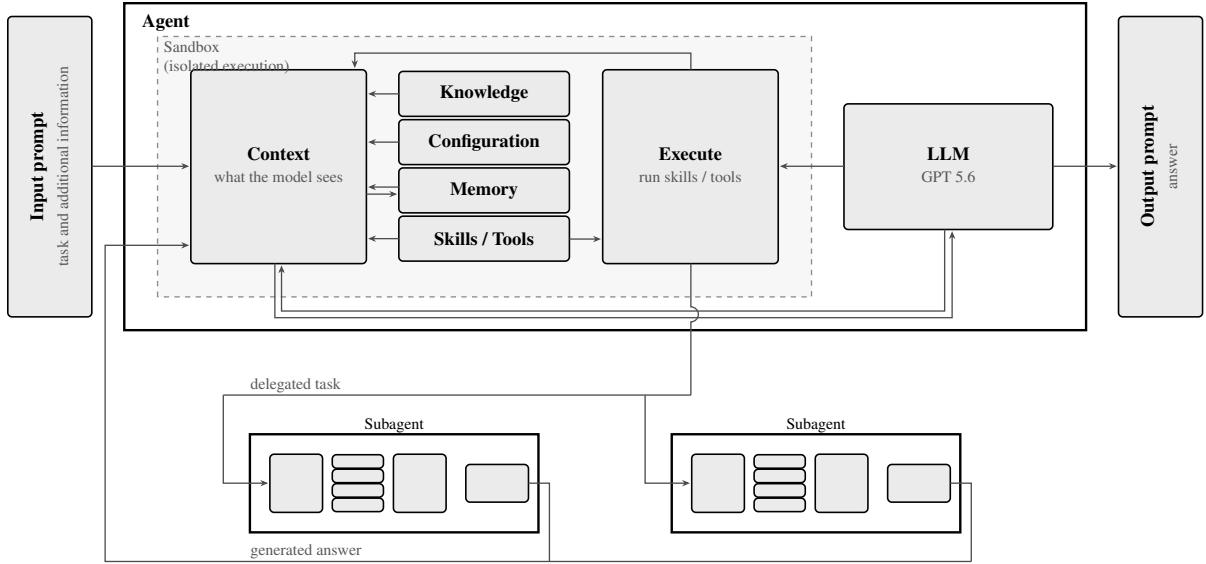

The harness elements of Section \ref{subsec:2:1} are configured as follows. The ReAct loop runs isolated inside a container
\citep{yaoReActSynergizingReasoning2023,openclaw}. The configuration files describe the available skills
and how they work together, and they assign each subagent its role and the subset of skills it may access. Knowledge is
provided as an ordinary file in the workspace that the agent reads only if it decides to, and no
retrieval module places it into the context \citep{guModelScalingSystem2026}. Memory is used within a session and forms a
second source of information next to the knowledge file, while memory across sessions is beyond the scope of this work.  
The two applications require different skills and knowledge, which the agentic system keeps apart. The system comprises
one orchestrating agent and multiple subagents, each responsible for a specific task. Along the axes described in
Section \ref{subsec:2:2}, communication is centralized in a star structure, collaboration is cooperative, and the
strategy is role-based. By design, the subagents do not communicate with one another. Each subagent receives a task from
the orchestrating agent and returns its result to it. Figure \ref{fig:agentic} illustrates the system.
The agent described in this section is applied to two tasks. Since they come from different areas of acoustics, 
they show the range of problems the approach can address. They also pose different questions about the agent. 
The first asks what each harness element contributes, and the second asks how the work is best divided between 
the skills while the other elements of the harness remain unchanged.

\emph{Fault detection in measured data.} In Section \ref{sec:3} the agent searches recordings for measurement faults.
Eight prompts ask for this in increasing detail. The simplest task asks whether a recording is faulty or clean, the
hardest task asks which of the nine fault types occurred and in which time interval. The configuration of the agent, 
the knowledge and the LLM are varied in isolation over the eight tasks and every answer is graded against a known
ground truth.

\emph{Design of an acoustic metamaterial.} In Section \ref{sec:4} the agent designs a ventilated metamaterial that
reduces the noise behind it over a given frequency range. The reduction can be reached with scatterers or with
resonators, and the agent chooses which and searches for the parameters that reach the target. How much of that work
is given as a skill changes between runs, while the rest of the harness stays fixed. The design is judged by the noise
reduction it achieves.
\section{\label{sec:3} Fault Detection in Synthetic Measurement Data}
Measurement campaigns for noise, vibration, and harshness (NVH) use numerous sensors in harsh environments, where
faults resulting from sensor failures and handling errors are prevalent \citep{baoComputerVisionDeep2019}. 
Nearly every analysis of the measurement data
is preceded by a check of the recordings for such faults. The methods range from a visual inspection of the recorded
channels to a script designed to detect the most common faults \citep{defreitasIdentificationAnomaliesMicrophone2023}.

This detection is the task the agent is given here. Section \ref{subsec:3:1} describes the synthetic recordings and the faults
injected into them, Section \ref{subsec:3:2} the skills the agent is equipped with, and Section \ref{subsec:3:3} the
configurations that are compared. Section \ref{subsec:3:4} defines how an answer is scored, and Sections
\ref{subsec:3:5} and \ref{subsec:3:6} report how the agent scores and what follows from it for practice.
\subsection{\label{subsec:3:1} The Benchmark Dataset}
The data for this benchmark are artificial recordings of interior noise in an electric vehicle, stored as stereo WAV files of ten
seconds at a sampling rate of $44.1\,\mathrm{kHz}$. Each clean sample is synthesized from source models of the vehicle, such as the
powertrain and the tires, and filtered with a cabin transfer function. \citep{dielSupervisedSourceSeparation2026}

Synthetic data offers two advantageous properties on which a benchmark depends. The first is an exact ground truth. The faults are
injected into a copy of the clean sample, so the type, time interval, and severity of every fault are recorded at the
moment of injection without later annotation. Every run is then graded against the same reference. The second is that
the content of the set is decided in advance. Every fault type appears, and the share of clean and corrupted recordings
is fixed, rather than being whatever a measurement campaign happens to deliver.

\begin{figure}
\resizebox{\textwidth}{!}{\input{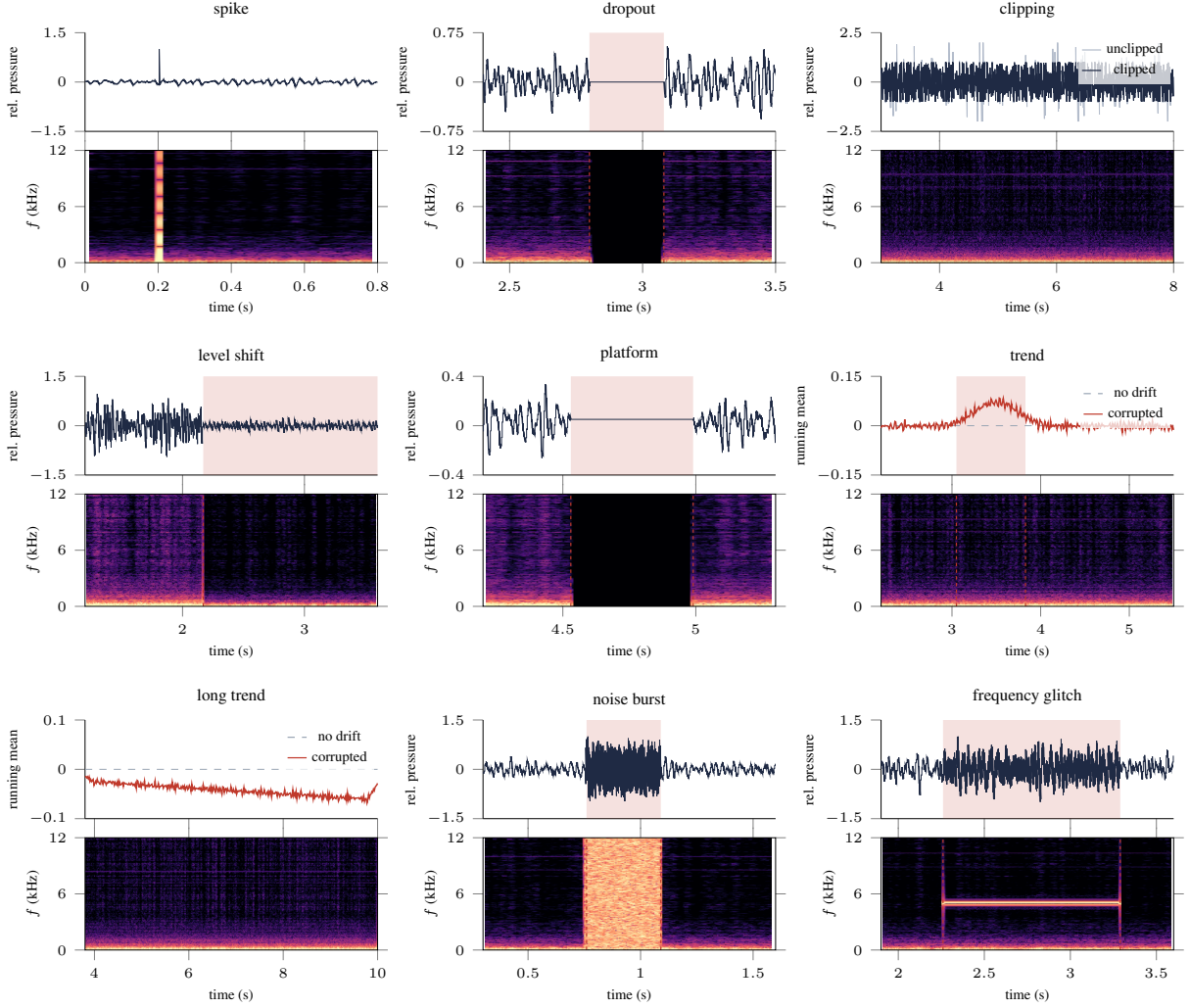}}
\caption{\label{fig:anomalies} The nine fault types of the fault detection task. Each panel shows the waveform or the
running mean of an affected recording at the top and the corresponding spectrogram below. Red dashed lines and the
shaded span mark the start and end of the fault. For clipping, the unclipped signal (gray) is drawn behind the clipped
one (blue). For trend and long trend, the running mean of the corrupted signal (red) is compared with a flat baseline
without drift (gray).}
\end{figure}

Nine fault types are injected, each representing a failure that can occur in a real measurement chain. 
Figure \ref{fig:anomalies} illustrates an example of each. The fault types spike, platform, dropout, trend, and long
trend are inspired by \citep{jiaValidationTimeSeries2024}.
The benchmark comprises $50$ curated recordings, of which $25$ are clean and $25$ are corrupted. With this balanced split, 
an agent that labels every recording as clean is correct in only half of the cases, so a score above this level reflects 
how well the agent distinguishes clean from corrupted recordings.
\subsection{\label{subsec:3:2} Skill Design}
The harness described in Section \ref{subsec:2:3} is generic. It is tailored to the task through the skills assigned to
the fault detection subagent that examines the recordings for faults. The subagent comprises $13$ skills, implemented as
Python functions and invoked through the execution tool. They form a two-stage pipeline in which four primitives measure
the recordings and nine detectors decide.

Each primitive skill transforms a recording into a feature artifact. One reduces the recording to a frame-by-frame
measure of root mean square (RMS), variance, and zero-crossing rate. A second provides a short-time spectrum with its
flatness and dominant frequency. A third tracks the amplitude at a given sampling resolution, and the fourth counts the
samples at full amplitude and the flat sections between them.

Each of the nine fault types has its own detector. A detector operates on an
artifact produced by a primitive. It compares the feature of that artifact with a
set of thresholds and returns a decision for its fault type, together with the time interval in which it detected the
fault. The agent takes the thresholds from the curated knowledge file. 
If no thresholds are provided, the detector fails instead of falling back to a default, which leaves the agent to
come up with a value of its own.

The split of the skills separates measurement from judgment. The primitives determine what the
agent perceives, the detectors what it concludes. Because the two stages are separate executables, the agent performs
the handover itself, running a primitive and passing the resulting artifact to the detector it considers appropriate.
How the skills are separated is a design choice of the engineer, because it fixes what is left for the agent to decide,
for example which primitive to run, which detector to apply, and how to connect the two. The more the skills fix, the more
reproducible the run and the fewer decisions the agent makes. The more is left open, the less certain the outcome, and
the more the record shows about how the agent works with its skills. The split used here serves the second purpose by 
making it possible to compare configurations by which skills they call and how reliably they call them.
\subsection{\label{subsec:3:3} Experimental Design}
The benchmark measures the contributions of the individual components of the harness to the performance of the agent. A
defined procedure and a clear definition of a measured run are prerequisites for this measurement. A run is defined as a
single session of the agent. In this session, the agent receives a prompt, works in its isolated container with the
skills and files of the configuration under test, and returns its response as a structured text block.
This structure allows the response of each run to be extracted and graded automatically. Each run starts in the same
workspace and is independent of previous runs, so two runs of a configuration differ only in what the agent itself does.
Memory is limited to a single session, and storage across sessions is beyond the scope of this work
\citep{luoStorageExperienceSurvey2026}. The response is archived and evaluated against the ground truth from Section
\ref{subsec:3:1}.

Each run is initiated with one of eight prompts. The
prompts are phrased in non-technical language and match the scenario the benchmark addresses, namely a user who asks
with limited prior knowledge of fault detection. The wording of a prompt is not a neutral entry point for the agent, as
it influences the skills the agent selects and the manner in which it evaluates the result
\citep{loyaExploringSensitivityLLMs2023,luFantasticallyOrderedPrompts2022,fangComprehensiveSurveySelfEvolving2025}. For
this reason, each prompt is fixed in its exact wording. 
\begin{table}
\caption{\label{tab:tasks}The task prompts of the benchmark. They vary along two axes: what is asked, rising from a
yes-or-no judgment (001, 002, 005) through the fault types present (003, 007) to the types with their time ranges
(004, 008), and how many recordings are given, one for 001 to 005 and a batch of $50$ recordings for 006 to 008.
Every prompt requests its answer as a structured text block.}
\centering
\small
\begin{tabular}{l p{0.8\textwidth}}
\toprule
No. & Input Prompt \\
\midrule
001 & Is anything wrong with this recording? (with a true positive sample) \\
002 & Is anything wrong with this recording? (with a true negative sample) \\
003 & Which faults are present in this recording and of which types? \\
004 & Which faults are present, of which types, and in which time ranges? \\
005 & A colleague is convinced that this recording is corrupted, identify the fault and its location. \\
006 & For each of these recordings, is it corrupted? \\
007 & For each of these recordings, which types of faults are present? \\
008 & For each of these recordings, which faults occur, of which type and in which time range? \\
\bottomrule
\end{tabular}
\end{table}
The benchmark follows a one-at-a-time analysis. Each phase changes a single element of the harness while the rest
stays fixed, so that a difference in the scores can be attributed to that element. The first phase determines which
parts of the harness an agent needs to solve the task. The second phase corrupts the information provided to the agent, and the third replaces the language model
underlying the harness. The first phase uses all eight prompts, and the latter two use a subset that covers the
different types of requests. Together, the three phases separate the contribution of the elements of the harness
from its robustness and its dependence on the model.

\begin{itemize}
\item \emph{Phase 1: Harness Elements.} The contribution of each element is measured by comparing four configurations. 
The configuration \emph{none} provides no skills, \emph{undocumented} provides
the skills without their documentation, \emph{baseline} adds the documentation, and \emph{knowledge} adds the thresholds 
as knowledge. Comparing none with undocumented separates the availability of a skill from its
description, and comparing baseline with knowledge shows the contribution of the thresholds and whether the agent reads 
its knowledge and applies it. 
\item \emph{Phase 2: Robustness of the Harness.} Robustness is tested by making the information available to the agent inconsistent. 
A corrupted memory entry contradicts the curated knowledge in the workspace, and the observation is
which of the two sources the agent follows. The literature refers to this as a knowledge conflict
\citep{xuKnowledgeConflictsLLMs2024,xieAdaptiveChameleonStubborn2024,longpreEntityBasedKnowledgeConflicts2022}.
\item \emph{Phase 3: Model Dependence.} The dependence on the model is measured by running the same harness with 
different language models.
\end{itemize}

Each combination of configuration and prompt is executed ten times, because an agent behaves probabilistically in two
ways. The language model samples its answer, and the loop turns each sampled decision into a branch, which increases the
number of possible paths with the length of a run
\citep{mohammadiEvaluationBenchmarkingLLM2025,luoLargeLanguageModel2025}. Two runs of the same configuration can follow
different paths and yield different answers, so a single run is one sample of a distribution. 
Ten runs are a modest number, but they offer insight into the reliability of a result, 
and distinguish a consistent result from one achieved only once.
\subsection{\label{subsec:3:4} Evaluation Metrics}
An agent benchmark can be graded by comparing the values reported by the agent with a reference or by having a language
model assess the responses. The latter approach is common \citep{wangMCPBenchBenchmarkingToolUsing2025}, but such a judge
is biased by construction \citep{liEvaluatingScoringBias2025}. In this work, a script compares the reported values with
the ground truth established in Section \ref{subsec:3:1}. This comparison requires a metric, a rule that reduces the
performance of one run to a single number. The prompts pose a challenge, as they request different kinds of responses,
from a binary verdict to the limits of a time interval. Each kind of response requires its own grading method for the
score to reflect the performance of the agent.

\begin{itemize}
\item \emph{Binary score.} A single recording that only has to be judged corrupted or clean is either right or wrong. A
run scores $0$ or $1$, and the result is reported as $k$ of $N$ runs correct. This metric is used for the
single-recording prompts listed in Table \ref{tab:grading}.
\item \emph{Balanced accuracy.} Conventional accuracy is calculated by dividing the number of correctly classified
recordings by the total number of recordings. In a set of $90$ clean and $10$ corrupted recordings, an agent that labels
every recording as clean achieves a score of $0.90$ without distinguishing clean from corrupted recordings
\citep{brodersenBalancedAccuracyIts2010}. A measure independent of this mix is the mean of the accuracies achieved on the
corrupted and on the clean recordings,
\begin{equation}
\text{balanced acc.} = \frac{1}{2}\left(\frac{TP}{TP+FN} + \frac{TN}{TN+FP}\right),
\end{equation}
which equals the conventional accuracy when the agent performs equally well on both classes and falls to chance level
for an agent that always gives the same answer. This metric is used for a batch of recordings, as shown in Table
\ref{tab:grading}.
\item \emph{$F_1$ score.} Answers that consist of a set of type names or fault locations are evaluated by the precision
$P$ and the recall $R$. The precision is the fraction of the named types that are actually present, and the recall is
the fraction of the present types that were named. For a recording containing a dropout, a spike, and clipping, an
answer naming a dropout and a level shift attains a precision of $1/2$, because one of the two named types is present,
and a recall of $1/3$, because one of the three present types was named. For the localization prompts, the same two
quantities are counted over events, and a prediction counts only if its type matches and its interval falls within a
tolerance given in Appendix \ref{app:matching}. With the reference set $Y$ and the answer set $\hat{Y}$, the two
quantities are given by
\begin{equation}
P = \frac{|Y \cap \hat{Y}|}{|\hat{Y}|}, \qquad R = \frac{|Y \cap \hat{Y}|}{|Y|}.
\end{equation}
A single number per cell is required, so the two are combined by the harmonic mean
\begin{equation}
F_1 = \frac{2\,P\,R}{P + R},
\end{equation}
which is dominated by the smaller of the two, so only an answer that is both complete and free of invented types
receives a high score \citep{vanrijsbergenInformationRetrieval1979}. Table \ref{tab:grading} lists the prompts graded in
this way.

One alternative for these prompts is the chance-corrected Matthews correlation coefficient, which also uses the true
negatives
\citep{matthewsComparisonPredictedObserved1975,chiccoAdvantagesMatthewsCorrelation2020,powersEvaluationPrecisionRecall2020}.
A true negative is well defined only for the binary decision, which is why
this work reports the $F_1$ score, following the convention of the literature on sound event detection
\citep{mesarosMetricsPolyphonicSound2016}.
\end{itemize}

Every prompt is reduced to a single value, which poses an additional challenge for prompts that require the analysis of
a batch of recordings. Such a batch can be reduced in two ways, by scoring each recording and averaging, or by pooling
the counts across the entire batch and reducing once. Pooling has the advantage that it weights each fault equally, not
each recording. A clean recording scored clean contributes nothing, a false alarm contributes one false positive, and a
recording the agent never answered contributes all of its faults as false negatives
\citep{sokolovaSystematicAnalysisPerformance2009,manningIntroductionInformationRetrieval2008}.

The scores are complemented by their spread across the repetitions and the resulting resolution. Without a measure of
spread, a difference between two configurations may lie within the noise, which would render the claim that one is
better unsubstantiated. How the spread is reported depends on the metric. For the balanced accuracy and the pooled
$F_1$, it is the standard deviation $\sigma$ and the resolution $\Delta$ over the ten runs. 
The resolution $\Delta$ is the smallest difference between two means that can be distinguished from the run-to-run noise,
\begin{equation}
\Delta \approx \frac{2\,\sigma}{\sqrt{N}}.
\end{equation}
A difference between two means has to exceed the resolution by approximately a
factor of $\sqrt{2}$ before it can be claimed \citep{agarwalDeepReinforcementLearning}. 

For the single-recording binary tasks, a run scores $0$
or $1$, so a spread would carry no information beyond the mean, and these tasks are given as $k$ of $N$ runs correct
\citep{brownIntervalEstimationBinomial2001}.
\begin{table}
\caption{\label{tab:grading}The six task groups, the metric applied to each, and whether the metric is pooled over a batch.}
\centering
\small
\begin{tabular}{l l l l}
\toprule
 & Prompts & Metric & Pooled \\
\midrule
\multicolumn{4}{l}{\emph{single recording}}\\
binary & 001, 002, 005 & binary score & no \\
type & 003 & $F_1$ & no \\
localization & 004 & $F_1$ & no \\
\midrule
\multicolumn{4}{l}{\emph{batch of recordings}}\\
binary & 006 & balanced accuracy & yes \\
type & 007 & $F_1$ & yes \\
localization & 008 & $F_1$ & yes \\
\bottomrule
\end{tabular}
\end{table}
\subsection{\label{subsec:3:5} Results}
This section reports the runs performed with the harness and the data described above, graded with the metrics
introduced before. Unless stated otherwise, all runs use the model GPT 5.6 terra. Several of the statements below are based on a
single run or on counts over the runs of a configuration, which the tables do not show. Each of these statements carries
the identifier of an entry in the supplementary material (e.g., E1), where the claim is set out together with the part
of the archived record it is based on.

As expected, adding harness elements clearly improved the performance of the agent, particularly the addition of skills
and of the knowledge file. Documentation on how to use the skills did not demonstrably improve the performance. Some
tasks scored higher and some lower, mostly within the noise.

The configuration none regularly underperforms the others, with one notable exception in the batch binary task. Tasks
001 and 002 ask the same question of a corrupted and of a clean recording, and together they give this configuration a
balanced accuracy of $0.60$, slightly above the one half reached by answering the same way for every recording. On the
batch of task 006, it attains $0.89$ and surpasses the baseline, because in a batch it ranks the recordings against one
another instead of against a threshold (E5). As expected, the score declines sharply when the tasks become more
difficult, for example when the fault has to be localized.

One notable observation is that undocumented performed significantly better on the clean cases 002 and 005. This cannot
be attributed to the missing documentation, because the performance of both undocumented and baseline depends strongly
on the thresholds they have to devise themselves. Within the ten recorded runs, undocumented may simply have chosen
better thresholds.

Both configurations reach moderate scores, which indicates that the invented thresholds are more than a guess. The
logs show where they come from. The agent runs the primitives, reads the magnitude of their output, and adjusts its
thresholds accordingly. When a pass reports no fault, the agent repeats it with lower thresholds until faults appear
(E3). In this way, it performs an iterative search that the task never asked for, and it stops once the result appears
plausible, without any cost function.

The undocumented and baseline configurations differ in the route they take through the skills. Baseline runs the
detectors in almost every run, while undocumented reaches for a primitive and then decides for itself. The documentation
thus changes which route the agent prefers, and the agent finds the skills in both cases. Table \ref{tab:skilluse} holds
the counts, and Table \ref{tab:cost} holds the number of skill calls, which differ by two orders of magnitude between
the two configurations. Both routes cost about the same. Baseline runs longer and spends more tokens on each batch task,
but no difference between the two configurations exceeds the combined resolution of the cells it is read from.

The fully equipped configuration scores perfectly in almost every run. Reading the knowledge file is optional, yet it
was consulted in every run, and the thresholds it provides reached the detectors unaltered. Two runs fall short, the
$0.95$ on task 006 and a $0.996$ on task 007, which Table \ref{tab:results} rounds to $1.00$. The first lost the results
it had recorded and reported every recording as anomalous (E1). The second submitted the remnants of an abandoned
attempt that remained in the workspace shared by the workers (E2). Neither failure can be attributed to skill handling.
\begin{table}
\newcommand{\rc}[3]{\makecell{#1\\[-1.5pt]{\scriptsize$\pm$#2\,(#3)}}}
\newcommand{\kn}[1]{\makecell{#1/10\\[-1.5pt]{\scriptsize\phantom{0}}}}
\caption{\label{tab:results}Results of the benchmark. Each cell reports the mean over $N = 10$ runs with the standard deviation and, in brackets,
the resolution beneath it. The single-recording binary tasks are given as the number of correct runs out of ten.}
\centering
\small
\begin{tabular}{l c c c c}
\toprule
Task & none & \makecell{undocu-\\mented} & baseline & knowledge \\
\midrule
\multicolumn{5}{l}{\emph{single recording}}\\
001 & \kn{\textbf{10}} & \kn{\textbf{10}} & \kn{\textbf{10}} & \kn{\textbf{10}} \\
002 & \kn{2} & \kn{5} & \kn{2} & \kn{\textbf{10}} \\
005 & \kn{1} & \kn{9} & \kn{3} & \kn{\textbf{10}} \\
003 & \rc{0.65}{0.28}{0.18} & \rc{0.73}{0.17}{0.11} & \rc{0.88}{0.16}{0.10} & \rc{\textbf{1.00}}{0.00}{0.00} \\
004 & \rc{0.26}{0.25}{0.16} & \rc{0.60}{0.16}{0.10} & \rc{0.69}{0.14}{0.09} & \rc{\textbf{1.00}}{0.00}{0.00} \\
\midrule
\multicolumn{5}{l}{\emph{batch of recordings}}\\
006 & \rc{0.89}{0.12}{0.07} & \rc{0.89}{0.21}{0.13} & \rc{0.74}{0.21}{0.13} & \rc{\textbf{0.95}}{0.16}{0.10} \\
007 & \rc{0.43}{0.17}{0.11} & \rc{0.61}{0.21}{0.13} & \rc{0.82}{0.11}{0.07} & \rc{\textbf{1.00}}{0.01}{0.01} \\
008 & \rc{0.29}{0.16}{0.10} & \rc{0.60}{0.13}{0.08} & \rc{0.69}{0.15}{0.09} & \rc{\textbf{1.00}}{0.00}{0.00} \\
\bottomrule
\end{tabular}
\end{table}
The next two phases use a reduced prompt set that covers the different types of requests, because the full set would
require considerably more tokens without providing additional information.
The next phase examines how the knowledge configuration handles inconsistent information. For this purpose, a corrupted
version of the threshold table is added to the memory file of the agent. Alternatively, the corrupted table could be
added to the knowledge as a second option. The setup also shows which source the agent prefers, given the instructions
in its configuration file. Two versions of these instructions were tested. In the first, both sources are mentioned
equally. In the second, the instructions strongly favor the knowledge by asking the agent to consult it for information
on thresholds, and only the last paragraph states that the memory is an alternative.
\begin{table}
\caption{\label{tab:sourcechoice}Which of the two contradicting threshold tables a run used, and under which instructions. Each row contains thirty runs,
ten for each of the tasks 004, 006, and 008. The three conditions are instructions that name both sources equally,
instructions that favor the knowledge over the memory, and the fair instructions with a raised reasoning effort.}
\centering
\small
\begin{tabular}{l c c c}
\toprule
Instructions & \makecell{curated\\only} & \makecell{corrupted\\only} & both \\
\midrule
fair & 25/30 & 4/30 & 1/30 \\
knowledge favored & 30/30 & 0/30 & 0/30 \\
fair, raised reasoning & 17/30 & 2/30 & 11/30 \\
\bottomrule
\end{tabular}
\end{table}
The agent does not divide its attention evenly between the two sources. Under fair instructions, it selected the curated
table in $25$ of the $30$ runs and the corrupted one in $4$. Where the instructions name the knowledge first, it selected the
curated table in all $30$ runs. The present setup cannot explain why the curated table is preferred under fair
instructions. A run that selected a table used this table exclusively, with one exception that ran the detectors twice, once with
each table, and answered from the curated set (E6).

The decisive factor is what the instructions advertise. Naming the knowledge first is sufficient to eliminate the
corrupted picks. The opposite direction was not tested.

Raising the reasoning effort changes which source the agent reaches for, not how well it answers. It consults both
sources in $11$ of the $30$ runs, against $1$ of $30$ at the lower setting, and the scores of the fair condition in Table \ref{tab:corruption} do not
move with it. Naming the knowledge first does move them. That condition reaches the values the knowledge configuration
attains without a conflict in Table \ref{tab:results}, so an instruction that points at the right source recovers
what the conflict costs, while a longer deliberation does not.

\begin{figure}
\resizebox{\textwidth}{!}{
\definecolor{mcTerra}{HTML}{4C72B0}
\definecolor{mcLuna}{HTML}{DD8452}
\definecolor{mcGrid}{HTML}{C9D4E6}
\begin{tikzpicture}
\begin{axis}[
  ybar, bar width=6.5pt,
  width=15.2cm, height=5.4cm,
  ymin=0, ymax=1.16, ytick={0,0.25,0.5,0.75,1},
  yticklabel style={font=\scriptsize},
  xmin=0.2, xmax=14.8,
  xtick={1,2,3,4,6,7,8,9,11,12,13,14},
  xticklabels={none,undoc.,baseline,knowl.,none,undoc.,baseline,knowl.,none,undoc.,baseline,knowl.},
  x tick label style={font=\scriptsize, rotate=40, anchor=east},
  ylabel={score}, ylabel style={font=\scriptsize},
  ymajorgrids, grid style={mcGrid, line width=0.4pt},
  axis line style={mcGrid, line width=0.6pt}, tick style={mcGrid},
  error bars/error bar style={black, line width=0.4pt},
  legend style={at={(1.02,1)}, anchor=north west, font=\scriptsize,
                draw=mcGrid, inner sep=2pt, row sep=-1pt},
  legend image code/.code={\draw[#1] (0cm,-0.08cm) rectangle (0.22cm,0.12cm);},
]
\addplot[fill=mcTerra, draw=mcTerra, error bars/.cd, y dir=both, y explicit] coordinates {
  (1,0.26) +- (0,0.16) (2,0.60) +- (0,0.10) (3,0.69) +- (0,0.09) (4,1.00) +- (0,0.00)
  (6,0.89) +- (0,0.07) (7,0.89) +- (0,0.13) (8,0.74) +- (0,0.13) (9,0.95) +- (0,0.10)
  (11,0.29) +- (0,0.10) (12,0.60) +- (0,0.08) (13,0.69) +- (0,0.09) (14,1.00) +- (0,0.00)};
\addplot[fill=mcLuna, draw=mcLuna, error bars/.cd, y dir=both, y explicit] coordinates {
  (1,0.20) +- (0,0.17) (2,0.49) +- (0,0.14) (3,0.35) +- (0,0.14) (4,0.80) +- (0,0.19)
  (6,0.77) +- (0,0.12) (7,0.85) +- (0,0.09) (8,0.69) +- (0,0.15) (9,0.94) +- (0,0.10)
  (11,0.14) +- (0,0.06) (12,0.26) +- (0,0.09) (13,0.54) +- (0,0.04) (14,0.80) +- (0,0.20)};
\legend{5.6 terra, 5.6 luna}
\node[anchor=south, font=\scriptsize] at (axis cs:2.5,1.02) {task 004};
\node[anchor=south, font=\scriptsize] at (axis cs:7.5,1.02) {task 006};
\node[anchor=south, font=\scriptsize] at (axis cs:12.5,1.02) {task 008};
\draw[mcGrid, line width=0.5pt] (axis cs:5,0) -- (axis cs:5,1.00);
\draw[mcGrid, line width=0.5pt] (axis cs:10,0) -- (axis cs:10,1.00);
\end{axis}
\end{tikzpicture}}
\caption{\label{fig:models}The four configurations on the tasks 004, 006, and 008. The color indicates the model that drives the agent, GPT 5.6
terra or GPT 5.6 luna. Each bar is the mean over ten runs, and the error bar is the resolution, so two bars with
overlapping error bars cannot be distinguished by this benchmark. On 006, no configuration separates the two models, on
004 they separate under baseline and knowledge, and on 008 the weaker model scores lower under every
configuration.}
\end{figure}
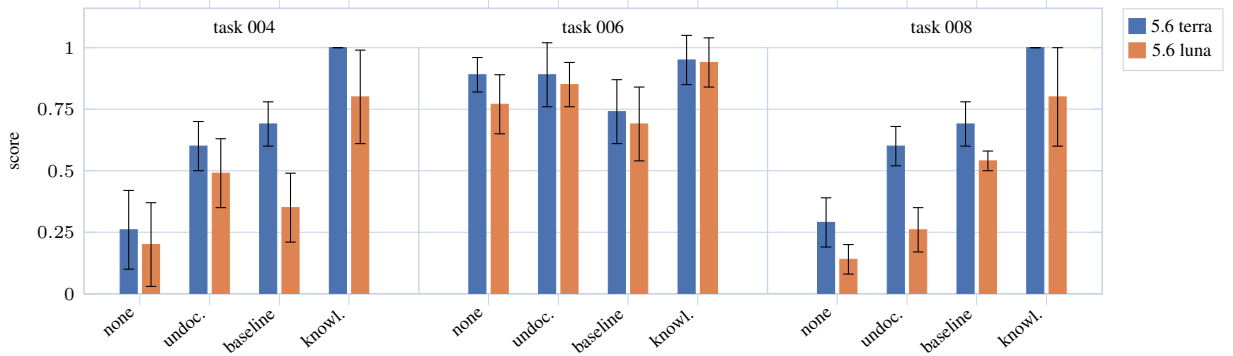

The final phase replaces the language model that drives the harness, GPT 5.6 terra, with GPT 5.6 luna and leaves the
harness, the recordings, and the prompts unchanged. Table \ref{tab:luna} reports the results, and Figure
\ref{fig:models} compares them to the previous results.
On the batch binary task 006, the scores of all configurations remain within the noise. On task 004, the result is
split. Baseline and knowledge score significantly lower on the weaker model, and the other two configurations remain
within the noise. On task 008, every configuration loses, and the fully equipped one drops from a perfect score to
$0.80$ with a standard deviation of about $0.3$, so its answers become unreliable. With both models, the fully equipped
configuration scores best on every task. The influence of the harness is thus visible regardless of the model that
drives it, and the ladder of Section \ref{subsec:3:3} survives the exchange. Table \ref{tab:lunacost} reports the cost
of a run on the weaker model.
\begin{table}
\newcommand{\rc}[3]{\makecell{#1\\[-1.5pt]{\scriptsize$\pm$#2\,(#3)}}}
\caption{\label{tab:luna}The four configurations on the weaker model GPT 5.6 luna. The harness, the recordings, and the prompts are those of
Table \ref{tab:results}, and only the model that drives the loop is exchanged. Each cell is the mean over $N = 10$ runs
with the standard deviation and, in brackets, the resolution beneath it.}
\centering
\small
\begin{tabular}{l c c c c}
\toprule
Task & none & \makecell{undocu-\\mented} & baseline & knowledge \\
\midrule
004 & \rc{0.20}{0.27}{0.17} & \rc{0.49}{0.22}{0.14} & \rc{0.35}{0.22}{0.14} & \rc{\textbf{0.80}}{0.31}{0.19} \\
006 & \rc{0.77}{0.18}{0.12} & \rc{0.85}{0.15}{0.09} & \rc{0.69}{0.23}{0.15} & \rc{\textbf{0.94}}{0.16}{0.10} \\
008 & \rc{0.14}{0.10}{0.06} & \rc{0.26}{0.14}{0.09} & \rc{0.54}{0.06}{0.04} & \rc{\textbf{0.80}}{0.32}{0.20} \\
\bottomrule
\end{tabular}
\end{table}
\subsection{\label{subsec:3:6} Discussion and Implications for the Practitioner}
The objective of the benchmark was to separate the contribution of the equipment of an agent from that of its model.

The benchmark runs demonstrate that the agent configuration defines the workflow the agent follows, although it cannot
determine in advance how well the agent will perform. If the skills are present but their execution is not described,
the agent writes its own code to use them. If the execution is described, the agent carries it out as intended. Once the
configuration is set, the workflow is predictable. This provides an agent developer with a method of debugging the
agent, since the steps of a run are known in advance and an error can be traced to the step that caused it.

The design of a configuration extends beyond the workflow of a single agent. It also determines how an agentic system
coordinates its work, which can be distributed among several agents, each with its own role and context. In this work,
the agent that coordinates the analysis is separate from the agent that performs the detection. Such a separation of
roles comes at a cost. Information has to be exchanged between the workspaces of the agents, and a handover can fail or
be misinterpreted by the receiving agent. An error of this kind is attributed to the orchestration, not to a skill or
its design. It cannot be eliminated completely, so an agent developer should account for it, for instance by repeating a
run.

A higher reasoning effort cannot compensate for structural errors or inconsistent information in the harness.
Orchestration is an integral part of the harness, and the harness determines which information reaches the model. If
this information is incomplete or contradictory, the cause lies in the structure of the system, and additional reasoning
by the model cannot resolve it. The agent in this work encounters two contradictory sources, and its responses show no
improvement with increased reasoning effort. An agent that performs poorly should be examined at the harness level
before its reasoning effort is increased, since a structural error persists regardless of the effort expended. The user
of an agent should also know which tasks the agent is equipped to handle. Reasoning alone does not guarantee that a task
outside this scope comes within reach, and the response of the agent is then likely to be hallucinated. Section
\ref{subsec:2:1} states that a harness cannot raise an agent beyond the limits imposed by its model. The results of this
work show the converse, namely that a model cannot compensate for structural problems of the harness.
A position paper about agentic AI scientists reaches the same conclusion, arguing that some challenges require
different design choices rather than more capable models \citep{bishtAgenticAIScientists2026}.

Finally, the environment of an agent may contain artifacts that were never intended for it. Data incorporated for a
specific purpose remains available, and an agent that finds it may react to it in ways no one anticipated. In this
benchmark, a run of the configuration without skills locates the retained skill files in the file system and executes
them, while its record reports no skill calls (E7). A benchmark has to account for the possibility that isolation fails,
and the output of an agent has to be traced back to its source before it can be trusted. A restricted environment is the
simplest measure, because what an agent cannot reach cannot affect it. Where the environment cannot be restricted
further, the log of a run shows the path taken and has to be reviewed before the score is accepted.
\section{\label{sec:4} Design of a Metamaterial with a Predefined Insertion Loss}
The second application shifts the focus from judging data to producing a design.
The agent is tasked with designing a periodic array representing a metamaterial 
that remains open to airflow and reduces the sound transmitted
through it at a given frequency \citep{krasikovaBroadbandNoiseinsulatingPeriodic2024}. Each task defines an insertion loss to be reached over a
frequency band, with restrictions on the maximum depth of the array, and the minimum clear opening shown in Figure \ref{fig:metasetup}. 
\subsection{\label{subsec:4:1} The Design Problem}
This section describes the physical model of the problem with its field equation and boundary conditions 
and defines the insertion loss.
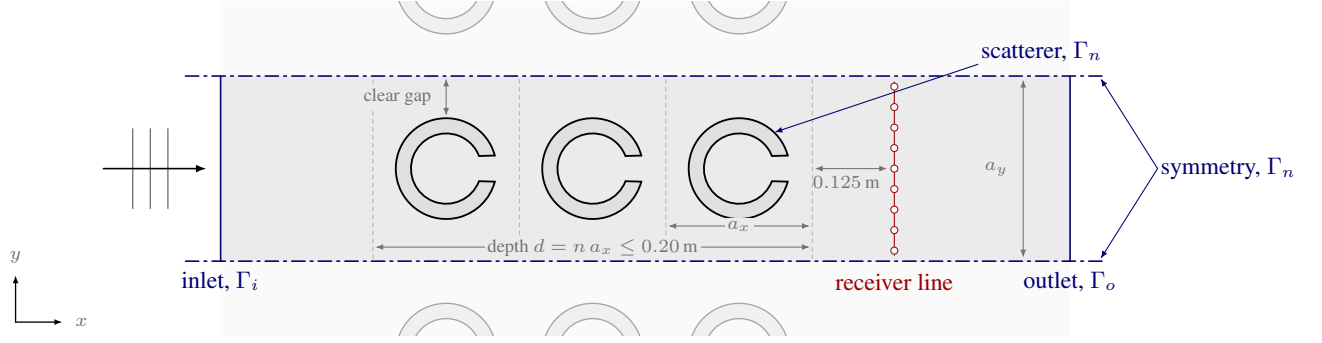
\begin{figure}
\centering
%
\begin{tikzpicture}[
  x=1.55cm, y=1.55cm,
  line cap=round, font=\footnotesize,
  sym/.style   = {draw=blue!45!black, line width=0.6pt, dash pattern=on 5pt off 2pt on 1pt off 2pt},
  port/.style  = {draw=blue!45!black, line width=0.6pt},
  ghost/.style = {draw=black!35, line width=0.5pt, fill=black!6},
  cell/.style  = {draw=black, line width=0.7pt, fill=black!12},
  dimen/.style = {draw=black!55, line width=0.4pt, {Latex[length=3pt]}-{Latex[length=3pt]}},
  ext/.style   = {draw=black!55, line width=0.3pt},
  lead/.style  = {draw=black!70, line width=0.4pt, -{Latex[length=3pt]}},
  lbl/.style   = {font=\footnotesize, text=black},
  dlbl/.style  = {font=\scriptsize, text=black!55},
]
\def\ax{1.25}      
\def\ay{1.58}      
\def\ncol{3}       
\def\xin{-1.30}    
\def\xout{5.95}    
\def\xrec{4.45}    
\def\Ro{0.43}      
\def\Ri{0.30}      
\def\ao{15}        
\def\ai{21}        

\newcommand{\cscat}[3]{
  \path[#3] ([shift={(\ao:\Ro)}]#1,#2) arc[start angle=\ao, end angle=360-\ao, radius=\Ro]
         -- ([shift={(-\ai:\Ri)}]#1,#2) arc[start angle=-\ai, end angle=-360+\ai, radius=\Ri]
         -- cycle;
}

\def\ghostshow{0.40}   
\begin{scope}
\clip (\xin,{-\ghostshow*\ay}) rectangle (\xout,{\ay+\ghostshow*\ay});
\foreach \row in {1,-1}{
  \fill[black!2] (\xin,{\row*\ay}) rectangle (\xout,{\row*\ay+\ay});
  \foreach \c in {0,1,2}{
    \cscat{\c*\ax+0.5*\ax}{\row*\ay+0.5*\ay}{ghost}
  }
}
\end{scope}

\fill[black!8] (\xin,0) rectangle (\xout,\ay);
\foreach \c in {0,1,2}{
  \cscat{\c*\ax+0.5*\ax}{0.5*\ay}{cell}
}
\foreach \c in {0,1,2,3}{
  \draw[draw=black!30, line width=0.3pt, dash pattern=on 1.5pt off 1.5pt]
    ({\c*\ax},0.02) -- ({\c*\ax},\ay-0.02);
}

\draw[sym] (\xin-0.30,0)   -- (\xout+0.30,0);
\draw[sym] (\xin-0.30,\ay) -- (\xout+0.30,\ay);

\draw[port] (\xin,0)  -- (\xin,\ay);
\draw[port] (\xout,0) -- (\xout,\ay);

\draw[draw=red!60!black, line width=0.4pt] (\xrec,0.05) -- (\xrec,\ay-0.05);
\foreach \i in {0,1,...,8}{
  \fill[white] (\xrec,{0.09+\i*(\ay-0.18)/8}) circle (0.030);
  \draw[draw=red!60!black, line width=0.35pt] (\xrec,{0.09+\i*(\ay-0.18)/8}) circle (0.030);
}

\foreach \x in {-2.05,-1.90,-1.75}{
  \draw[draw=black!55, line width=0.5pt] (\x,{0.5*\ay-0.34}) -- (\x,{0.5*\ay+0.34});
}
\draw[draw=black, line width=0.6pt, -{Latex[length=4pt]}] (-2.30,0.5*\ay) -- (-1.42,0.5*\ay);

\draw[dimen] (\xout-0.40,0.03) -- (\xout-0.40,\ay-0.03);
\node[dlbl, anchor=east] at (\xout-0.45,0.5*\ay) {$a_y$};

\draw[dimen] ({2*\ax+0.02},0.30) -- ({3*\ax-0.02},0.30);
\node[dlbl, fill=black!8, inner sep=1pt] at (2.5*\ax,0.30) {$a_x$};

\draw[dimen] (0.02,0.12) -- ({\ncol*\ax-0.02},0.12);
\node[dlbl, fill=black!8, inner sep=1pt]
  at ({0.5*\ncol*\ax},0.12) {depth $d = n\,a_x \le 0.20\,$m};

\draw[dimen] ({\ncol*\ax+0.02},0.5*\ay) -- ({\xrec-0.06},0.5*\ay);
\node[dlbl, anchor=north, inner sep=1pt] at ({0.5*(\ncol*\ax+\xrec)-0.06},{0.5*\ay-0.04}) {$0.125\,$m};

\draw[dimen] ({0.5*\ax},{0.5*\ay+\Ro+0.02}) -- ({0.5*\ax},{\ay-0.02});
\node[dlbl, anchor=east, fill=black!8, inner sep=1pt]
  at ({0.5*\ax-0.10},{0.5*(0.5*\ay+\Ro+\ay)}) {clear gap};

\node[lbl, text=blue!45!black, align=center, inner sep=1pt] (symbc)
  at ({\xout+1.35},{0.5*\ay}) {symmetry, $\Gamma_n$};
\draw[lead, draw=blue!45!black] (symbc.west) -- ({\xout+0.26},{\ay-0.02});
\draw[lead, draw=blue!45!black] (symbc.west) -- ({\xout+0.26},{0.02});

\node[lbl, anchor=south east, inner sep=1pt, text=blue!45!black] (scbc)
  at (\xout+0.30,\ay+0.10) {scatterer, $\Gamma_n$};
\draw[lead, draw=blue!45!black] (scbc.south west) -- ({2.5*\ax+\Ro*0.70},{0.5*\ay+\Ro*0.72});

\node[lbl, anchor=north, align=center, inner sep=1pt, text=blue!45!black]
  at (\xin,-0.06) {inlet, $\Gamma_i$};
\node[lbl, anchor=north, align=center, inner sep=1pt, text=blue!45!black]
  at (\xout,-0.06) {outlet, $\Gamma_o$};
\node[lbl, anchor=north, inner sep=2pt, text=red!60!black] at (\xrec,-0.06) {receiver line};

\draw[draw=black, line width=0.5pt, -{Latex[length=3pt]}] (\xin-1.75,-0.52) -- (\xin-1.35,-0.52);
\draw[draw=black, line width=0.5pt, -{Latex[length=3pt]}] (\xin-1.75,-0.52) -- (\xin-1.75,-0.12);
\node[dlbl, anchor=west] at (\xin-1.32,-0.52) {$x$};
\node[dlbl, anchor=south] at (\xin-1.75,-0.09) {$y$};
\end{tikzpicture}
\caption{\label{fig:metasetup}The simulated setup with scatterers or resonators in air, where the domain is governed by
the Helmholtz equation. The array repeats infinitely in the $y$-direction, and a plane wave travels in the $x$-direction. The wave enters at the inlet on the left
and leaves at the outlet on the right, and both ports absorb what reaches them. Every design respects a maximum
thickness of the metamaterial and a minimum gap between two unit cells. The insertion loss is evaluated on the
receiver line behind the array.}
\end{figure}
The field is computed in two dimensions in the frequency domain, where the acoustic pressure $p$ satisfies the
Helmholtz equation
\begin{equation}
\nabla^2 p + k^2 p = 0 ,
\end{equation}
with the wave number $k = 2\pi f / c$ and the speed of sound $c = 343\,\mathrm{m/s}$ in air.
The duct walls and the scatterer surfaces $\Gamma_n$ are rigid \citep{pierceAcousticsIntroductionIts2019},
and the inlet $\Gamma_i$ and the outlet $\Gamma_o$ absorb what reaches them through a
first-order absorbing condition \citep{engquistAbsorbingBoundaryConditions1977, baylissBoundaryConditionsNumerical1982},
\begin{align}
\partial_n p &= 0 && \text{on } \Gamma_n,\\
\partial_n p + \mathrm{i} k p &= 2\mathrm{i}k && \text{on } \Gamma_i,\\
\partial_n p + \mathrm{i} k p &= 0 && \text{on } \Gamma_o .
\end{align}
A plane wave of unit amplitude is injected at normal incidence through the inlet condition, whose right-hand side
follows from splitting the field into the incident and the reflected wave. The wave reflected back toward the inlet is
absorbed at the same time, so that no energy re-enters the domain from the boundary.
Figure \ref{fig:metasetup} shows the simulated domain with the three boundary conditions, the unit
cell and the receiver line behind the array. The simulation is performed in Python using the dolfinx package
\citep{barattaDOLFINxNextGeneration2023}. 
The weak form is discretized with quadratic Lagrange elements on an unstructured triangular mesh
\citep{thompsonReviewFiniteelementMethods2006} and solved directly at each frequency. The element size $h$ of the mesh is bounded by two rules, and the smaller of the two bounds applies,
\begin{equation}
h \le \min\!\left( \frac{c}{10\,f_{\mathrm{max}}},\; \alpha\, s_{\mathrm{min}} \right) .
\end{equation}
The first is a resolution rule for the Helmholtz equation, where $f_{\mathrm{max}}$ denotes the highest frequency of the band. 
No fixed number of elements per wavelength is sufficient in general, because the error also grows with the
number of wavelengths across the domain \citep{ihlenburgFiniteElementAnalysis1998,marburgPollutionEffectBoundary2018}.
Recommendations therefore vary with the element order and the problem \citep{marburgDiscretizationRequirementsHow2008},
and ten quadratic elements per shortest wavelength is a conservative choice for a domain that spans at most a few
wavelengths.
The second ties the element size to the smallest dimension of the cell $s_{\mathrm{min}}$, such as the slit width or the wall 
thickness, so that at least $1/\alpha$ elements span it \citep{krasikovaAcousticBoundStates2024}. This resolves the slit, whose width and length set the resonance frequency
\citep{ingardTheoryDesignAcoustic1953,selametHelmholtzResonatorExtended2003,elfordMatryoshkaLocallyResonant2011}. 
In the present designs, the second rule is decisive by a wide margin. At $1\,\mathrm{kHz}$,
the first rule allows an element of about $34\,\mathrm{mm}$, and the features of these cells reduce it to between
$1\,\mathrm{mm}$ and $5\,\mathrm{mm}$.

Only a single duct is simulated (see Figure \ref{fig:metasetup}). For a cell that is symmetric about its mid-plane and
a wave at normal incidence, the lateral rigid walls act as symmetry planes. The duct thus represents the whole
metamaterial, which repeats infinitely in the $y$ direction 
\citep{gebrekidanAutonomousDesignNoisemitigating2024,kronowetterSoundAttenuationEnhancement2023}.
This assumption holds only as long as the duct supports a single propagating mode. A symmetric cell at normal incidence excites only the even modes, so the
second mode appears first, and the duct remains single-mode below $f_c = c / a_y$ 
\citep{pierceAcousticsIntroductionIts2019}. Above $f_c$, the inlet and the outlet no
longer absorb exactly what reaches them. For this reason, the solver rejects any frequency above $0.8 f_c$, which keeps a margin 
below the limit. This is also the range over which the supplementary material compares the solver with the semi-analytic 
solution of Linton and Evans \citep{lintonInteractionWavesRow1993}, which applies under the same single-mode condition.

With the model defined and its physical limits established, a criterion for the evaluation is required. A design is
evaluated by its insertion loss on a receiver line behind the array \citep{krasikovaBroadbandNoiseinsulatingPeriodic2024}. Since a
wave of unit amplitude is injected at the inlet, the value in the empty channel is exactly $|p| = 1$. 
The insertion loss follows from the root mean square pressure alone 
\citep{munjalAcousticsDuctsMufflers2014, jenaInsertionLossRegular2021}, 
\begin{equation}
\mathrm{IL}(f) = -10 \log_{10} \big\langle |p|^2 \big\rangle_\mathrm{receiver} ,
\end{equation}
where a positive value indicates attenuation. The receiver line consists of $21$ points that are evenly distributed
across the channel and located at a fixed distance behind the material, as indicated in Figure \ref{fig:metasetup}.
\subsection{\label{subsec:4:2} Study Setup}
This case study analyzes the division between what a skill performs and what the agent has to work out for itself.
Section \ref{subsec:3:2} identified this division as the central decision in designing an agent. The model, the harness,
the targets, and the budget remain constant in every run, and only the set of skills changes. The three phases that are 
discussed below describe the different skill setups that are used. 
\begin{figure}
\centering
%
\begin{tikzpicture}[
  x=1.68cm, y=1.68cm, line cap=round, font=\footnotesize,
  solid/.style   = {draw=black, line width=0.5pt, fill=black!12},
  cellbox/.style = {draw=black!45, line width=0.35pt, dash pattern=on 2pt off 1.6pt},
  dimen/.style   = {draw=black, line width=0.3pt, {Latex[length=2.2pt]}-{Latex[length=2.2pt]}},
  lead/.style    = {draw=black, line width=0.3pt, -{Latex[length=2.2pt]}},
  tick/.style    = {draw=black, line width=0.3pt},
  plbl/.style    = {font=\footnotesize},
  dlbl/.style    = {font=\scriptsize, inner sep=0.8pt},
]
\def\axc{2.00}    
\def\ayc{1.60}    
\def\dx{2.42}     
\def\dy{2.05}     
\def\cy{0.80}     

\newcommand{\cshapeT}[5]{%
  \pgfmathsetmacro{\Rin}{#3-#4}%
  \pgfmathsetmacro{\aout}{asin((#5/2)/#3)}%
  \pgfmathsetmacro{\ain}{asin((#5/2)/\Rin)}%
  \path[solid] ([shift={(\aout:#3)}]#1,#2) arc[start angle=\aout, end angle=360-\aout, radius=#3]
            -- ([shift={(-\ain:\Rin)}]#1,#2) arc[start angle=-\ain, end angle=-360+\ain, radius=\Rin]
            -- cycle;
}

\newcommand{\wallthk}[8]{
  \pgfmathsetmacro{\Rinn}{#3-#4}%
  \pgfmathsetmacro{\rot}{(#5>90 && #5<270) ? #5-180 : #5}%
  \pgfmathsetmacro{\inx}{min(0.15,0.5*\Rinn)}%
  \draw[tick] ([shift={(#5:\Rinn-\inx)}]#1,#2) -- ([shift={(#5:#3+#6)}]#1,#2);
  \draw[lead] ([shift={(#5:#3+#6)}]#1,#2) -- ([shift={(#5:#3-0.005)}]#1,#2);
  \draw[lead] ([shift={(#5:\Rinn-\inx)}]#1,#2) -- ([shift={(#5:\Rinn+0.005)}]#1,#2);
  \node[dlbl, anchor=#7, inner sep=1.6pt] at ([shift={(#5:#3+#6)}]#1,#2) {#8};
}

\newcommand{\members}{%
  \draw[tick] (0,-0.05) -- (0,-0.50);
  \draw[tick] ({0.25*\axc},-0.05) -- ({0.25*\axc},-0.22);
  \draw[tick] ({0.75*\axc},-0.05) -- ({0.75*\axc},-0.50);
  \draw[dimen] (0.02,-0.16) -- ({0.25*\axc-0.02},-0.16);
  \node[dlbl, anchor=west] at ({0.25*\axc+0.05},-0.16) {$a_x/4$};
  \draw[dimen] (0.02,-0.42) -- ({0.75*\axc-0.02},-0.42);
  \node[dlbl, fill=white] at ({0.375*\axc},-0.42) {$3a_x/4$};
}

\begin{scope}[shift={(0,0)}]
  \draw[cellbox] (0,0) rectangle (\axc,\ayc);
  \draw[solid] ({0.5*\axc},\cy) circle (0.56);
  \draw[lead] ({0.5*\axc},\cy) -- ++(42:0.56);
  \node[dlbl, anchor=south west] at ([shift={(42:0.62)}]{0.5*\axc},\cy) {$r$};
  \node[plbl, anchor=south west] at (0,{\ayc+0.05}) {(a)};
\end{scope}

\begin{scope}[shift={(\dx,0)}]
  \def\Ro{0.60}\def\wl{0.15}\def\sl{0.34}
  \draw[cellbox] (0,0) rectangle (\axc,\ayc);
  \cshapeT{0.5*\axc}{\cy}{\Ro}{\wl}{\sl}
  \draw[lead] ({0.5*\axc},\cy) -- ++(112:\Ro);
  \node[dlbl, fill=white] at ([shift={(131:0.58*\Ro)}]{0.5*\axc},\cy) {$R$};
  \wallthk{0.5*\axc}{\cy}{\Ro}{\wl}{300}{0.20}{west}{$t$}
  \draw[dimen] ({0.5*\axc+\Ro-0.5*\wl},{\cy-0.5*\sl}) -- ({0.5*\axc+\Ro-0.5*\wl},{\cy+0.5*\sl});
  \node[dlbl, anchor=west] at ({0.5*\axc+\Ro-0.5*\wl+0.06},\cy) {$w$};
  \draw[dimen] ({\axc+0.20},0.02) -- ({\axc+0.20},{\ayc-0.02});
  \node[dlbl, anchor=west] at ({\axc+0.24},{0.5*\ayc}) {$a_y$};
  \node[plbl, anchor=south west] at (0,{\ayc+0.05}) {(b)};
\end{scope}

\begin{scope}[shift={(0,-\dy)}]
  \def\ra{0.42}\def\rb{0.28}
  \draw[cellbox] (0,0) rectangle (\axc,\ayc);
  \draw[solid] ({0.25*\axc},\cy) circle (\ra);
  \draw[solid] ({0.75*\axc},\cy) circle (\rb);
  \draw[lead] ({0.25*\axc},\cy) -- ++(125:\ra);
  \node[dlbl, anchor=south] at ([shift={(125:\ra+0.06)}]{0.25*\axc},\cy) {$r_1$};
  \draw[lead] ({0.75*\axc},\cy) -- ++(55:\rb);
  \node[dlbl, anchor=south] at ([shift={(55:\rb+0.06)}]{0.75*\axc},\cy) {$r_2$};
  \members
  \draw[dimen] (0.02,-0.70) -- ({\axc-0.02},-0.70);
  \node[dlbl, anchor=north] at ({0.5*\axc},-0.74) {$a_x$};
  \node[plbl, anchor=south west] at (0,{\ayc+0.05}) {(c)};
\end{scope}

\begin{scope}[shift={(\dx,-\dy)}]
  \def\Ra{0.44}\def\Rb{0.22}
  \def\ta{0.12}\def\tb{0.07}
  \draw[cellbox] (0,0) rectangle (\axc,\ayc);
  \cshapeT{0.25*\axc}{\cy}{\Ra}{\ta}{0.24}
  \cshapeT{0.75*\axc}{\cy}{\Rb}{\tb}{0.13}
  \draw[lead] ({0.25*\axc},\cy) -- ++(125:\Ra);
  \node[dlbl, anchor=south] at ([shift={(125:\Ra+0.06)}]{0.25*\axc},\cy) {$R_1$};
  \draw[lead] ({0.75*\axc},\cy) -- ++(125:\Rb);
  \node[dlbl, anchor=south] at ([shift={(125:\Rb+0.06)}]{0.75*\axc},\cy) {$R_2$};
  \draw[dimen] ({0.25*\axc+\Ra-0.5*\ta},{\cy-0.12}) -- ({0.25*\axc+\Ra-0.5*\ta},{\cy+0.12});
  \node[dlbl, anchor=west] at ({0.25*\axc+\Ra-0.5*\ta+0.05},\cy) {$w_1$};
  \draw[dimen] ({0.75*\axc+\Rb-0.5*\tb},{\cy-0.065}) -- ({0.75*\axc+\Rb-0.5*\tb},{\cy+0.065});
  \node[dlbl, anchor=west] at ({0.75*\axc+\Rb-0.5*\tb+0.05},\cy) {$w_2$};
  \wallthk{0.25*\axc}{\cy}{\Ra}{\ta}{300}{0.26}{west}{$t_1$}
  \wallthk{0.75*\axc}{\cy}{\Rb}{\tb}{300}{0.24}{west}{$t_2$}
  \members
  \draw[dimen] (0.02,-0.70) -- ({\axc-0.02},-0.70);
  \node[dlbl, anchor=north] at ({0.5*\axc},-0.74) {$a_x$};
  \draw[dimen] ({\axc+0.20},0.02) -- ({\axc+0.20},{\ayc-0.02});
  \node[dlbl, anchor=west] at ({\axc+0.24},{0.5*\ayc}) {$a_y$};
  \node[plbl, anchor=south west] at (0,{\ayc+0.05}) {(d)};
\end{scope}
\end{tikzpicture}
\caption{\label{fig:metatemplates}The four unit-cell templates. A rigid circle (a) scatters the incident wave, and a
c-shape (b), a cavity behind a narrow slit, resonates. The diatomic variants (c) and (d) contain two members of
different sizes in one cell, placed at one quarter and three quarters of the lattice spacing. The agent sets the lattice
spacings $a_x$ and $a_y$, the number of columns, and the shape parameters. These are the radius $r$ of a circle and the
outer radius $R$, the wall thickness $t$, and the slit width $w$ of a c-shape. Each member of a diatomic cell has its
own values.}
\end{figure}
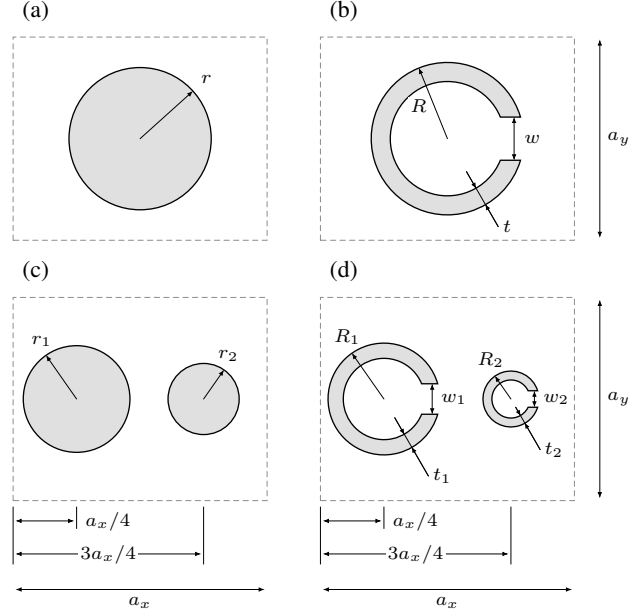
\begin{itemize}
\item \emph{Phase 1: Agent-driven optimization.} The agent is given the unit cell templates of Figure
\ref{fig:metatemplates} and skills that check the geometric constraints shown in Figure \ref{fig:metasetup}, 
mesh a cell and solve it. It chooses the type of unit cell and its
parameters, decides which parameter to change after each evaluation, and decides when to stop the optimization. 
The element size is decided by the agent as well, following the rule in the knowledge file.

\item \emph{Phase 2: Agent-orchestrated optimization.} The same skills as in the previous phase, plus four search strategies, 
implemented in Python with scipy \citep{2020SciPy-NMeth}, from a simplex search to uniform random sampling.
One of them takes the parameter search over: the agent picks which parameters to
vary, their bounds and the optimization budget, and the strategy returns the best design it found. 
The element size is now set by the skill. The strategy improves a design at one frequency only, while the target has to be met across 
the whole frequency band, so the agent has to steer the optimization across the band itself.

\item \emph{Phase 3: Agent-driven design.} The skills of the first phase, without the unit cell templates. 
The agent draws and meshes its own cell, and a check skill admits it only if it is symmetric about its center plane and 
the solver can read the mesh.
\end{itemize}

The first phase is run in a narrowband and a broadband version, which together with the second and the third phase
gives the four tasks listed in Table \ref{tab:targets}. Each task specifies an insertion loss that has to be reached at every frequency of the band given in
the table. Each task comes at three levels of difficulty. The easy bar lies within what a rigid scatterer can reach. The
middle and the hard bar lie above the easy bar and can be met only with a resonator.
\begin{table}
\caption{\label{tab:targets}The four design tasks and the insertion loss each has to reach. Task 101 asks for a narrow
band around $1\,\mathrm{kHz}$, and the others ask for a wider band around the same center. Task 103 adds the search
strategies, and task 104 withholds the templates. The insertion loss is required at every frequency of the band.}
\centering
\small
\begin{tabular}{l l c c c c}
\toprule
Task & Asked for & easy & middle & hard & \makecell{band\\(Hz)} \\
\midrule
101 & narrowband & $8\,\mathrm{dB}$ & $20\,\mathrm{dB}$ & $45\,\mathrm{dB}$ & 970--1030 \\
102 & broadband & $6\,\mathrm{dB}$ & $18\,\mathrm{dB}$ & $40\,\mathrm{dB}$ & 891--1122 \\
103 & broadband with search & $6\,\mathrm{dB}$ & $18\,\mathrm{dB}$ & $40\,\mathrm{dB}$ & 891--1122 \\
104 & broadband, no templates & $6\,\mathrm{dB}$ & $18\,\mathrm{dB}$ & $40\,\mathrm{dB}$ & 891--1122 \\
\bottomrule
\end{tabular}
\end{table}

After each run the grader checks the geometric limits, recomputes the design with half the element size, and
accepts the result given by the agent only if the insertion loss remains unchanged. The margin indicates by how much the lowest insertion
loss in the band exceeds the threshold. It is positive for a design that meets its target and negative for one that
falls short. Computing the same margin from the numbers reported by the agent shows whether the agent was right about
what it had built.
\subsection{\label{subsec:4:3} Results}
Table \ref{tab:metaresults} lists the twelve runs, one for each of the four tasks at each of the three levels of
difficulty. Eight runs reached their target, and all twelve remained within the parameter limits. The values the agent
reports for its own design agree with the independent evaluation. The largest deviation is $1.4\,\mathrm{dB}$, and in every
run, the agent's assessment of whether the target was reached matched that of the grader.

The cell column records the shape on which each run settled. At the middle and the hard level of difficulty, all eight
runs end with a resonator. At the easy level, where the bar is within reach of a scatterer, the results split two to two,
with resonators in tasks 101 and 102 and scatterers in tasks 103 and 104. 
Task 104 provides no template, and the shapes drawn there follow the same pattern, from a solid block for the easy
bar to cavities for the higher bars.

The design evaluation column counts how many candidates each run passed to the solver, out of the $60$ allowed by the
task. The runs of tasks 101, 102, and 104 used between $3$ and $10$ evaluations each. Only task 103, which includes the
search strategies, used more, with $20$, $46$, and $11$. Every run thus ended with a large part of the budget unused,
including the runs that stopped just short of their goal. The easy cell of 104 was submitted as a miss by half a
decibel with $53$ evaluations still remaining.
\begin{table}
\caption{\label{tab:metaresults}The results of the four tasks, each at three levels of difficulty. The cell column
gives the template the agent selected, and the layers column gives how many cells deep it made the array. The margin
column gives the smallest insertion loss over the graded band relative to the required value, so a negative margin is a
miss. The claim column gives the difference between the insertion loss the agent reported and the one found by an
independent rerun of the same design. The design evaluation column counts the calls of the solver. In task 104, the
agent submits a mesh and no template parameters, so the rerun refines this mesh and does not build a new mesh, unlike in the other cases. 
The claims listed there are the smallest in the table. Eight of the twelve designs reach the required insertion loss, and
claimed and graded values differ by no more than $1.4\,\mathrm{dB}$. Given templates, the agent selects the c-shape in
seven of the nine runs. Without them, it draws cells of its own, which are named after their function.}
\centering
\small
\begin{tabular}{l l r r r r}
\toprule
Level & Cell & layers & \makecell{margin\\(dB)} & \makecell{design\\eval.} & \makecell{claim\\(dB)} \\
\midrule
\multicolumn{6}{l}{\emph{101 tonal, solver without search}}\\
easy & c-shape & 2 & 3.5 & 4 & $-0.1$ \\
middle & c-shape & 2 & 10.9 & 7 & $-0.1$ \\
hard & c-shape & 2 & 2.6 & 5 & $-0.1$ \\
\midrule
\multicolumn{6}{l}{\emph{102 broadband, solver without search}}\\
easy & c-shape & 4 & 17.6 & 3 & $-0.3$ \\
middle & c-shape & 4 & 10.0 & 10 & $-1.4$ \\
hard & bicshape & 2 & $-27.0$ & 9 & $-0.2$ \\
\midrule
\multicolumn{6}{l}{\emph{103 broadband, search strategies given}}\\
easy & circle & 2 & 1.4 & 20 & $-0.1$ \\
middle & c-shape & 2 & $-0.9$ & 46 & $-0.4$ \\
hard & c-shape & 4 & 2.0 & 11 & $-1.1$ \\
\midrule
\multicolumn{6}{l}{\emph{104 broadband, templates withheld}}\\
easy & block & 2 & $-0.5$ & 7 & $0.0$ \\
middle & two cavities & 2 & 2.1 & 5 & $0.1$ \\
hard & one cavity & 2 & $-37.2$ & 6 & $0.0$ \\
\bottomrule
\end{tabular}
\end{table}

\begin{figure}
\centering
\input{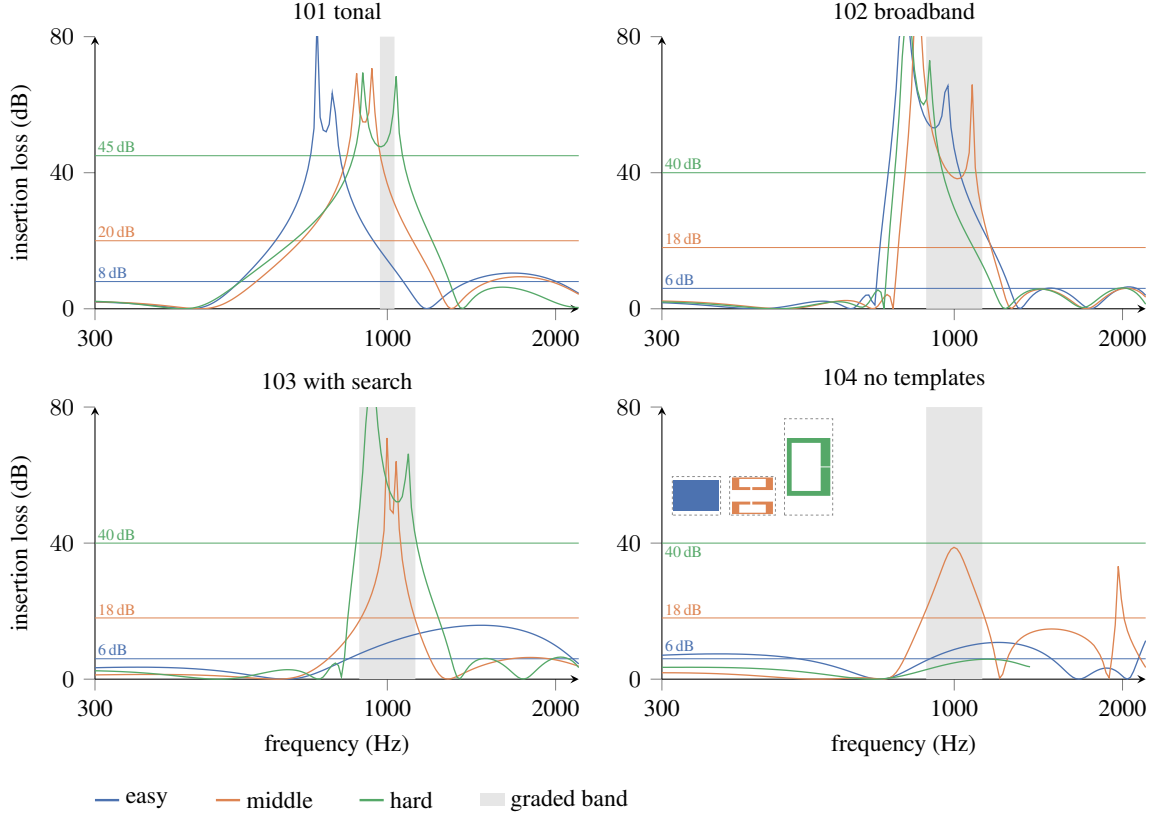}
\caption{\label{fig:metawide}Each submitted design, solved again independently over the range from $300$ to $2200$
hertz. The thin horizontal lines mark the insertion loss required at each level of difficulty, which is $6$, $18$, and
$40\,\mathrm{dB}$ in the broadband tasks and $8$, $20$, and $45\,\mathrm{dB}$ in 101. The gray stripe is the graded band.
Eight of the twelve designs remain above their line across that band. The middle level of 103 and the easy level of 104
miss it narrowly, and the hard levels of 102 and 104 miss it by a wide margin. The hard curve of 104 ends at $1.4\,\mathrm{kHz}$, because above this frequency its cell no longer satisfies the single-mode condition of Section
\ref{subsec:4:1}. The plot is cut at $80\,\mathrm{dB}$, and since the model has no losses, the height of the peaks has no
physical meaning.}
\end{figure}

Figure \ref{fig:metawide} recomputes each design far outside its target band and shows where its resonance was placed.
In tasks 101 and 102, where no search strategy is available, the peak of all six resonator designs lies below the band
by $30$ to $220\,\mathrm{Hz}$, so the bar is cleared on the flank of the resonance. In task 103, where the
strategies are available, both resonators reach their peak within the band. The third design there is a scatterer and
has no resonance to place. The strategies are responsible for placing the response at the desired location. A single run per cell makes this a pattern and not a
measurement, but the pattern is consistent across all designs in the comparison.

\begin{figure}
\centering
\resizebox{\textwidth}{!}{\input{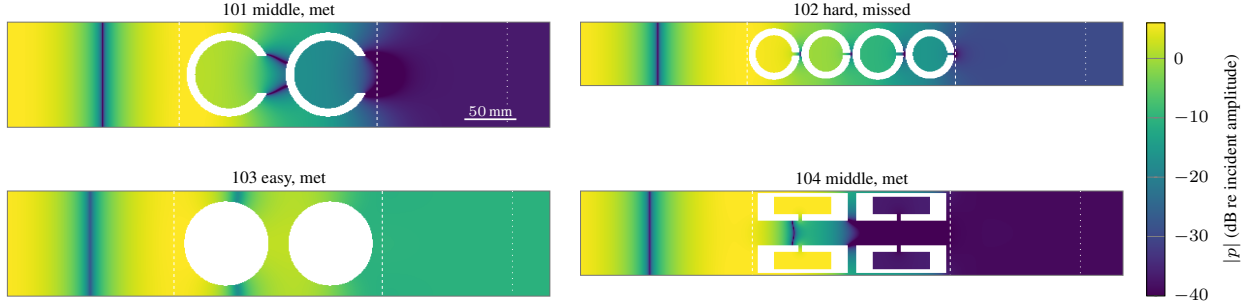}}
\caption{\label{fig:metafield}The pressure field of one design per task at $1\,\mathrm{kHz}$, plotted on a common length
scale, so that a taller cell is also drawn taller, and on a common color scale, so that the panels can be compared. In
each panel, the two dashed lines mark the front and the back of the array, the dotted line further to the right is the
receiver at which the insertion loss is read, and the white shapes are the scatterers. The dark vertical band in front
of each array is the first pressure node of the standing wave that the reflection builds up there, where the incident
and the reflected wave cancel each other. The panel of 102 shows a diatomic cell whose two members are so close in size
that its two layers appear as four. In 104 middle, the pair of cavities facing the wave is loud, and the pair behind it
is quiet. This is precisely the function of a resonator, and the agent drew it without a template to copy.}
\end{figure}

Figure \ref{fig:metafield} shows the pressure field of one design per task at $1\,\mathrm{kHz}$. All four fields share the
same color scale and can be compared directly. The dashed lines mark the front and the back of the array, which in these
designs lie at or just inside the depth limit of $0.20\,\mathrm{m}$. The dotted line behind them marks the receiver line at
which the insertion loss is measured. The color behind the array indicates how much each design attenuates at this
frequency. The three resonators reach $30$ to $39\,\mathrm{dB}$, and the single scatterer of 103 reaches about $11$. The
cell of 102 is of particular interest. It was built from the \emph{bicshape} template, which holds two members of
different size, yet its two members are nearly identical, with outer radii that differ by only $4\,\%$.
\subsection{\label{subsec:4:4} Discussion and Implications for the Practitioner}
The designs reported here are not advances in metamaterial design. The templates are standard cell types, the model is
two dimensional and without losses, and each design is the result of a single run. What this case study contributes is
the record of how the agent arrives at a design and of where the division between skill and agent decides the outcome.
The metamaterial is the testbed of this study, and the agent is its subject.

This case study produces a working agent for the design of an acoustic metamaterial. The physical model and the
simulation setup are provided, and the part that has to be adjusted is the geometry of the unit cell. The agent
optimized a given set of parameters on its own. Where a search strategy was available, the agent coordinated the use of
the strategy and selected which parameters to vary and what to optimize for. When no template was provided, it developed its own design within the
physical constraints of the model and optimized it. 
Every design it submitted satisfied the given constraints, and most reached their target.

The performance of the agent also has limits. It ends the design optimization prematurely. In most runs, fewer than $10$
of the $60$ allowed evaluations were used, and one design was submitted half a decibel below the threshold with $53$
evaluations still remaining. The task does not state that the full budget should be used, so the agent has no reason to use it.
When a run stops short of both the target and the budget, what it lacks is a skill or a rule that defines when to stop.
A more capable model would not fill this gap.

Without a search strategy, in tasks 101 and 102, the resonance of each design lies below
the graded band, so the
threshold is crossed on the flank of the peak. In addition, the agent favors safety over efficiency when choosing the
element size and uses a much finer mesh than the two rules of Section \ref{subsec:4:1} require.

Both the strengths and the weaknesses of the agent stem from the decisions it has to make around the skills. Skills can
thus be designed toward different goals, and the goal is determined by what a skill takes care of and what it leaves
open. If the answer to a design decision is already known, the decision is best fixed within a skill. The search
strategies are an example. Delegating the parameter search to a strategy tunes the resonator to the target band.
Without such a strategy, the cell is tuned to a neighboring range and still clears the bar within the graded band. The
strategies placed the attenuation where the goal required it. 
The interface between a design and the solver is fixed for the same reason, so the agent has to produce a
mesh that passes a check and does not have to infer the format the solver needs.

Where the designer does not know the answer, it is better to leave the decision open, because this is where exploration
by the agent has its place. In the final phase, no cell template was provided, and the agent generated its own cells.
This is useful whenever the form of a good design is not known in advance.
\section{\label{sec:5} Conclusion and Future Work}
This work treated the harness of an agent as the variable of an experiment. A benchmark on fault detection in synthetic
recordings of vehicle interior noise varied the elements of the harness, and a case study on the design of a periodic
array varied the division of work between a skill and the agent.
 
The elements of the harness differ in what they change. Skills and a curated set of thresholds change what the agent
achieves. A description of the workflow changes the route the agent takes without changing its accuracy, so that a run
becomes predictable and a fault can be traced to the step that caused it. Exchanging the model raises or lowers the
level at which all configurations settle, and the fully equipped configuration scores best with either model. A higher
reasoning effort changes which source of information the agent consults, while the quality of its answers remains the
same. The harness thus determines how an agent works, and the model determines how far it can get.
 
The design of a skill is a decision that the engineer has to make deliberately. Where a workflow is already established,
fixing it inside a skill costs nothing, as the search strategies and the solver interface show. Where no such workflow
exists, leaving the decision to the agent lets it explore designs that a script or template could not include, as the
unit cells the agent drew without a given example show. In general, designing a skill pipeline entails a trade-off
between reliability and exploration.
 
The study of language-model agents in acoustics is still at an early stage. More sophisticated agent designs, more demanding
applications, and complete simulation workflows remain to be examined. To retain control
over such systems, it will be essential to measure the contribution of each new element and how the elements work
together. This work can serve as a starting point for isolating the elements of an agent and measuring their impact.

\section*{Acknowledgements}
This work is a result of the research project ML/AI Support for NVH Engineering (FVV project no. 1555), funded by 
the FVV e.V. The project was carried out jointly by the Chair of Thermodynamics of Mobile Energy Conversion Systems 
at RWTH Aachen University, the Chair of Vibroacoustics of Vehicles and Machines at the Technical University of Munich, 
and the Data Analytics and Machine Learning group at Forschungszentrum Jülich. A group of experts led by 
Dr. Michael Fischer (Robert Bosch GmbH) accompanied the project. The authors gratefully acknowledge the support of the 
project coordination and of all members of the project committee.

\section*{Supplementary Material}
See the supplementary material for the scripts that re-derive every table and figure of this paper from the archived
records, the rules that map a reported fault type onto one of the nine types of the benchmark, the evidence behind the
statements of Section \ref{subsec:3:5}, cited above as E1 to E7, the four design prompts of Section \ref{sec:4} word
for word, and the validation of the solver against the semi-analytic solution of Linton and Evans
\citep{lintonInteractionWavesRow1993}.

\section*{Data and Code Availability}
The data that support the findings are part of the supplementary material. Specifications about the implementation of the
skills are available from the corresponding author upon reasonable request.

\section*{Author Declarations}
\subsection*{Use of Generative AI}
During the preparation of this work, the authors used Claude (Anthropic) and GPT 5.5 (OpenAI) to assist with coding, 
with version control in git, and with the LaTeX formatting of the manuscript, including the conversion of figures 
from Inkscape to TikZ. The authors also used these services to screen raw data and to improve the language of the 
manuscript, and used DeepL for further language editing. After using these services, the authors reviewed and edited 
the content as needed and take full responsibility for the content of the published article.

\subsection*{Conflict of Interest}
The authors declare no conflicts of interest regarding this manuscript. All authors contributed to the research and
writing of this paper.
 
\appendix

\section{Matching Reported Events}
\label{app:matching}

The agent names the faults it finds in its own words. Those names and the names in the ground truth are both
mapped onto the nine fault types of the benchmark before anything is counted, by the ordered rules given in the
supplementary material.

Where a prompt asks when a fault occurs, both the response and the reference are lists of type, onset and offset
\citep{mesarosMetricsPolyphonicSound2016}. For each reference event, the grader considers the unmatched predictions
of the same type and selects the one whose onset is closest. No optimal assignment is computed
\citep{kuhnHungarianMethodAssignment1955}. Each reference therefore absorbs at most one prediction, and a duplicate
detection becomes a false positive \citep{everinghamPascalVisualObject2010}. A match is counted if onset and offset
are both within a tolerance
\begin{equation}
\mathrm{tol.} = \min\Big(\max\big(0.05\,d,\ 0.10\,\mathrm{s},\ f\big),\ \max\big(0.25\,\mathrm{s},\ f\big)\Big),
\end{equation}
where $d$ is the duration of the reference event and $f$ is a floor determined by the window and the hop of the
primitive from which the detector reads. The floor is $0.60\,\mathrm{s}$ for trend and long trend, $0.15\,\mathrm{s}$ for level shift
and $0.10\,\mathrm{s}$ for the remaining types. It prevents a fault from being graded more finely than the instrument that
measures it, and the cap prevents a fault that persists throughout the entire recording from being graded as loosely
as its duration would allow. The criterion is a tolerance at the endpoints rather than a fraction of overlap
\citep{bilenFrameworkRobustEvaluation2020}, so it is binary per event and independent of how long a fault happens to
be.

\section{Additional Benchmark Results}
\label{app:cost}

This appendix holds what the archive records about a run, alongside what the run achieved in Table
\ref{tab:results}. The tables follow the phases of Section \ref{sec:3}: the configuration ladder, then the context
conflict and the model exchange.
Each cost is given as a mean with its standard deviation and resolution, as the scores are. In several cells the
standard deviation exceeds the mean, because a single run took far longer or spent far more than the others and
dominates the cell. These cells are skewed and should be read with caution.
\begin{table}
\newcommand{\cc}[3]{\makecell{#1\\[-1.5pt]{\scriptsize$\pm$#2\,(#3)}}}
\caption{\label{tab:cost}Cost of a run for the four configurations of Phase 1, on GPT 5.6 terra. Each cell is the mean over $N = 10$ runs, with the
standard deviation and, in brackets, the resolution beneath it, as in Table \ref{tab:results}. Tokens are thousands of tokens per
run and the wall clock time is seconds per run, both summed over the main agent and its workers.}
\centering
\small
\begin{tabular}{l c c c c}
\toprule
Task & none & \makecell{undocu-\\mented} & baseline & knowledge \\
\midrule
\multicolumn{5}{l}{\emph{skill calls}}\\
001 & \cc{0}{0}{0} & \cc{2}{1}{1} & \cc{10}{4}{2} & \cc{\textbf{14}}{2}{1} \\
002 & \cc{0}{0}{0} & \cc{3}{2}{1} & \cc{13}{2}{2} & \cc{\textbf{14}}{1}{1} \\
005 & \cc{0}{0}{0} & \cc{5}{2}{1} & \cc{\textbf{16}}{4}{2} & \cc{14}{1}{0} \\
003 & \cc{0}{0}{0} & \cc{3}{1}{1} & \cc{13}{0}{0} & \cc{\textbf{14}}{1}{0} \\
004 & \cc{0}{0}{0} & \cc{2}{2}{1} & \cc{13}{1}{1} & \cc{\textbf{16}}{5}{3} \\
006 & \cc{0}{0}{0} & \cc{8}{10}{6} & \cc{746}{225}{142} & \cc{\textbf{828}}{237}{150} \\
007 & \cc{0}{0}{0} & \cc{26}{61}{39} & \cc{\textbf{1041}}{506}{320} & \cc{1034}{462}{292} \\
008 & \cc{0}{0}{0} & \cc{8}{19}{12} & \cc{965}{359}{227} & \cc{\textbf{1228}}{199}{126} \\
\midrule
\multicolumn{5}{l}{\emph{thousands of tokens}}\\
001 & \cc{45}{8}{5} & \cc{79}{39}{24} & \cc{83}{28}{18} & \cc{\textbf{96}}{41}{26} \\
002 & \cc{64}{23}{14} & \cc{96}{41}{26} & \cc{92}{16}{10} & \cc{\textbf{101}}{28}{18} \\
005 & \cc{99}{42}{27} & \cc{\textbf{142}}{58}{36} & \cc{124}{40}{25} & \cc{98}{28}{18} \\
003 & \cc{76}{15}{10} & \cc{79}{26}{17} & \cc{\textbf{93}}{25}{16} & \cc{84}{18}{11} \\
004 & \cc{74}{22}{14} & \cc{\textbf{109}}{46}{29} & \cc{93}{22}{14} & \cc{90}{28}{18} \\
006 & \cc{\textbf{394}}{406}{257} & \cc{315}{113}{71} & \cc{346}{157}{100} & \cc{329}{107}{68} \\
007 & \cc{298}{108}{68} & \cc{281}{149}{94} & \cc{\textbf{794}}{1225}{775} & \cc{360}{122}{77} \\
008 & \cc{366}{130}{82} & \cc{435}{148}{94} & \cc{582}{284}{180} & \cc{\textbf{672}}{386}{244} \\
\midrule
\multicolumn{5}{l}{\emph{seconds}}\\
001 & \cc{90}{12}{8} & \cc{110}{41}{26} & \cc{96}{37}{24} & \cc{\textbf{119}}{71}{45} \\
002 & \cc{105}{14}{9} & \cc{\textbf{120}}{22}{14} & \cc{106}{27}{17} & \cc{105}{21}{13} \\
005 & \cc{127}{90}{57} & \cc{\textbf{196}}{110}{69} & \cc{155}{41}{26} & \cc{118}{67}{43} \\
003 & \cc{\textbf{129}}{57}{36} & \cc{127}{48}{30} & \cc{113}{51}{32} & \cc{119}{54}{34} \\
004 & \cc{105}{35}{22} & \cc{115}{31}{20} & \cc{\textbf{132}}{46}{29} & \cc{92}{16}{10} \\
006 & \cc{332}{178}{112} & \cc{306}{99}{62} & \cc{394}{149}{94} & \cc{\textbf{409}}{134}{85} \\
007 & \cc{304}{75}{48} & \cc{308}{112}{71} & \cc{\textbf{735}}{725}{459} & \cc{453}{108}{68} \\
008 & \cc{358}{109}{69} & \cc{460}{198}{125} & \cc{\textbf{676}}{237}{150} & \cc{635}{314}{199} \\
\bottomrule
\end{tabular}
\end{table}
Table \ref{tab:skilluse} aggregates which skills have been used in the eighty runs for each agent configuration. 
\begin{table}
\caption{\label{tab:skilluse}Which skills each configuration reached for, over the eighty runs of a configuration. Every
cell is counted from the reports the skills wrote. A run counts as \emph{primitives only} when it produced a
feature artifact and never called a detector, so its decisions were made somewhere other than in the detectors.}
\centering
\small
\begin{tabular}{l c c c}
\toprule
Configuration & \makecell{any skill\\report} & \makecell{primitives\\only} & \makecell{all nine\\detectors} \\
\midrule
none & 1/80 & 1/80 & 0/80 \\
undocumented & 59/80 & 29/80 & 0/80 \\
baseline & 80/80 & 0/80 & 71/80 \\
knowledge & 80/80 & 0/80 & 79/80 \\
\bottomrule
\end{tabular}
\end{table}
Table \ref{tab:corruption} provides the performance scores of the three context corruption conditions. 
Most of the performance drops stem from the corrupted tables that were selected by the agent over the correct table. 
How often each table was selected is reported in Table \ref{tab:sourcechoice}.
\begin{table}
\newcommand{\kc}[3]{\makecell{#1\\[-1.5pt]{\scriptsize$\pm$#2\,(#3)}}}
\caption{\label{tab:corruption}The scores of the three conditions in which the curated thresholds were contradicted
by corrupted ones in the memory of the agent. Each cell is the mean over $N = 10$ runs, with the standard deviation
and, in brackets, the resolution beneath it, as in Table \ref{tab:results}, where the same tasks are reported
without a conflict. The third condition repeats the fair one with the reasoning effort of the model
raised.}
\centering
\small
\begin{tabular}{l c c c}
\toprule
Task & fair & \makecell{knowledge\\favored} & \makecell{fair, raised\\reasoning} \\
\midrule
004 & \kc{0.94}{0.19}{0.12} & \kc{1.00}{0.00}{0.00} & \kc{0.88}{0.25}{0.16} \\
006 & \kc{0.98}{0.06}{0.04} & \kc{1.00}{0.00}{0.00} & \kc{0.98}{0.05}{0.03} \\
008 & \kc{0.90}{0.21}{0.13} & \kc{0.99}{0.02}{0.01} & \kc{0.99}{0.03}{0.02} \\
\bottomrule
\end{tabular}
\end{table}
Table \ref{tab:lunacost} shows the same pattern of skill calls, time spent and tokens used as on terra in Table \ref{tab:cost}.
\begin{table}
\newcommand{\lc}[3]{\makecell{#1\\[-1.5pt]{\scriptsize$\pm$#2\,(#3)}}}
\caption{\label{tab:lunacost}What a run costs on GPT 5.6 luna, the weaker model, to be read against Table
\ref{tab:cost}. Each
cell is the mean over $N = 10$ runs, with the standard deviation and, in brackets, the resolution beneath it.}
\centering
\small
\begin{tabular}{l c c c c}
\toprule
Task & none & \makecell{undocu-\\mented} & baseline & knowledge \\
\midrule
\multicolumn{5}{l}{\emph{skill calls}}\\
004 & \lc{0}{0}{0} & \lc{2}{2}{1} & \lc{13}{0}{0} & \lc{\textbf{14}}{2}{1} \\
006 & \lc{0}{0}{0} & \lc{1}{3}{2} & \lc{1033}{1160}{733} & \lc{\textbf{1334}}{994}{629} \\
008 & \lc{0}{0}{0} & \lc{0}{0}{0} & \lc{1104}{826}{523} & \lc{\textbf{1146}}{775}{490} \\
\midrule
\multicolumn{5}{l}{\emph{thousands of tokens}}\\
004 & \lc{59}{19}{12} & \lc{74}{26}{17} & \lc{\textbf{76}}{24}{15} & \lc{53}{9}{6} \\
006 & \lc{106}{58}{37} & \lc{143}{143}{90} & \lc{326}{121}{77} & \lc{\textbf{349}}{79}{50} \\
008 & \lc{185}{122}{77} & \lc{210}{71}{45} & \lc{318}{69}{44} & \lc{\textbf{438}}{125}{79} \\
\midrule
\multicolumn{5}{l}{\emph{seconds}}\\
004 & \lc{\textbf{107}}{8}{5} & \lc{102}{21}{13} & \lc{96}{28}{18} & \lc{103}{17}{11} \\
006 & \lc{111}{38}{24} & \lc{144}{105}{66} & \lc{438}{297}{188} & \lc{\textbf{551}}{354}{224} \\
008 & \lc{158}{77}{49} & \lc{144}{33}{21} & \lc{388}{148}{94} & \lc{\textbf{570}}{303}{192} \\
\bottomrule
\end{tabular}
\end{table}

\clearpage
\bibliographystyle{unsrtnat}
\bibliography{references}

\end{document}